\documentclass[
reprint,
superscriptaddress,
aps,
amsmath,amssymb,
longbibliography,
]{revtex4-2}

\usepackage{graphicx}%
\usepackage{dcolumn}%
\usepackage{bm}%
\usepackage{xspace}
\usepackage{xcolor}
\usepackage{hyperref}

\newcommand{\be}{\begin{equation}}
\newcommand{\ee}{\end{equation}}
\newcommand{\bal}{\begin{align}}
\newcommand{\eal}{\end{align}}

\newcommand{\MoI}{MoI$_3$\xspace}

\newcommand{\cv}{{\bf c}}
\newcommand{\gv}{{\bf g}}

\newcommand{\Av}{{\bf A}}
\newcommand{\Bv}{{\bf B}}
\newcommand{\Hv}{{\bf H}}

\newcommand{\Uv}{{\bf U}}

\newcommand{\Dv}{{\bf D}}

\newcommand{\Sv}{{\bf S}}
\newcommand{\Sg}{{\Sigma}}

\newcommand{\Xv}{{\bf X}}
\newcommand{\tv}{{\bf t}}

\newcommand{\Iv}{{\bf I}}
\newcommand{\Gv}{{\bf G}}
\newcommand{\w}{\omega}
\newcommand{\hw}{\hbar\omega}

\newcommand{\xh}{{\hat{x}}}
\newcommand{\yh}{{\hat{y}}}

\newcommand{\nh}{{\hat{n}}}

\newcommand{\muv}{{\boldsymbol \mu}}

\newcommand{\ajd}{{a_j^\dagger}}

\newcommand{\tJl}{{\tilde{J}_1}}
\newcommand{\tJtwo}{{\tilde{J}_2}}

\begin{document}

\title{Topological Magnonic Properties of an Antiferromagnetic Chain} %

\author{Topojit Debnath}
\email{tdebn001@ucr.edu}
\affiliation{Laboratory for Terahertz $\&$ Terascale Electronics (LATTE), Department of Electrical and Computer Engineering, University of California, Riverside, CA, 92521, USA}

\author{Shri Hari Soundararaj}
\email{ssoun005@ucr.edu}
\affiliation{Materials Science and Engineering, University of California, Riverside, CA, 92521, USA}

\author{Sohee Kwon}
\email{skwon054@ucr.edu}
\affiliation{Laboratory for Terahertz $\&$ Terascale Electronics (LATTE), Department of Electrical and Computer Engineering, University of California, Riverside, CA, 92521, USA}

\author{Alexander A. Balandin}
\email{balandin@seas.ucla.edu}
\affiliation{Department of Materials Science and Engineering, University of California, Los Angeles, CA 90095, USA}

\author{Roger K. Lake}
\email[Corresponding author: ]{rlake@ucr.edu}
\affiliation{Laboratory for Terahertz $\&$ Terascale Electronics (LATTE), Department of Electrical and Computer Engineering, University of California, Riverside, CA, 92521, USA}

\date{\today}%

\begin{abstract}
The magnonic excitations of a dimerized, one-dimensional, antiferromagnetic chain 
can be trivial or topological depending on the signs and magnitudes of the 
alternating exchange couplings and the anisotropy.
The topological phase that occurs when the signs of the two different
exchange couplings alternate 
is qualitatively different from that of the  Su-Schrieffer-Heeger  model.
A material that may exhibit these properties is
the quasi-one-dimensional material MoI$_3$ that
consists of dimerized chains weakly 
coupled to adjacent chains. 
The magnetic ground state and its excitations are analyzed both analytically 
and numerically using exchange and anisotropy parameters extracted 
from density functional theory calculations. 
\end{abstract}

\maketitle

\section{Introduction}
\label{sec:Intro}
Anti-ferromagnetism and topological magnonics 
\cite{2018_Topo_Magnonics_Paradigm_PRApp,
2021_Topo_Magnon_Review_PhysRep,
2022_Topo_Magnons_review_McClarty,
2022_AFM_Nernst_Cheng_APL}
in low-dimensional materials are of high current interest.
The majority of the work in topological magnonics has focused on 
two-dimensional (2D) ferromagnetic (FM) and antiferromagnetic (AFM) systems. 
The 2D systems generally rely on the presence of the
Dzyaloshinskii-Moriya interaction, which serves as an analogue to
spin-orbit coupling in electronic systems, to obtain a topological magnonic 
phase.
In one-dimensional electronic systems, dimerization can result in 
a topological phase described by the Su-Schrieffer-Heeger (SSH) model
\cite{1979_SSH_PRL},
and in one-dimensional FM systems, dimerization results in
a magnetic analogue of the SSH model 
with a topological magnonic phase \cite{2018_SSH_Magnon_Chain,2022_SSH_Magnons_JPCM}.
We will show that dimerization alone does not qualitatively alter the magnonic
spectrum in an AFM spin chain, and that more is required to obtain a
topological magnonic phase.

\begin{figure}[t]
\centering
\includegraphics[width=3.2in]{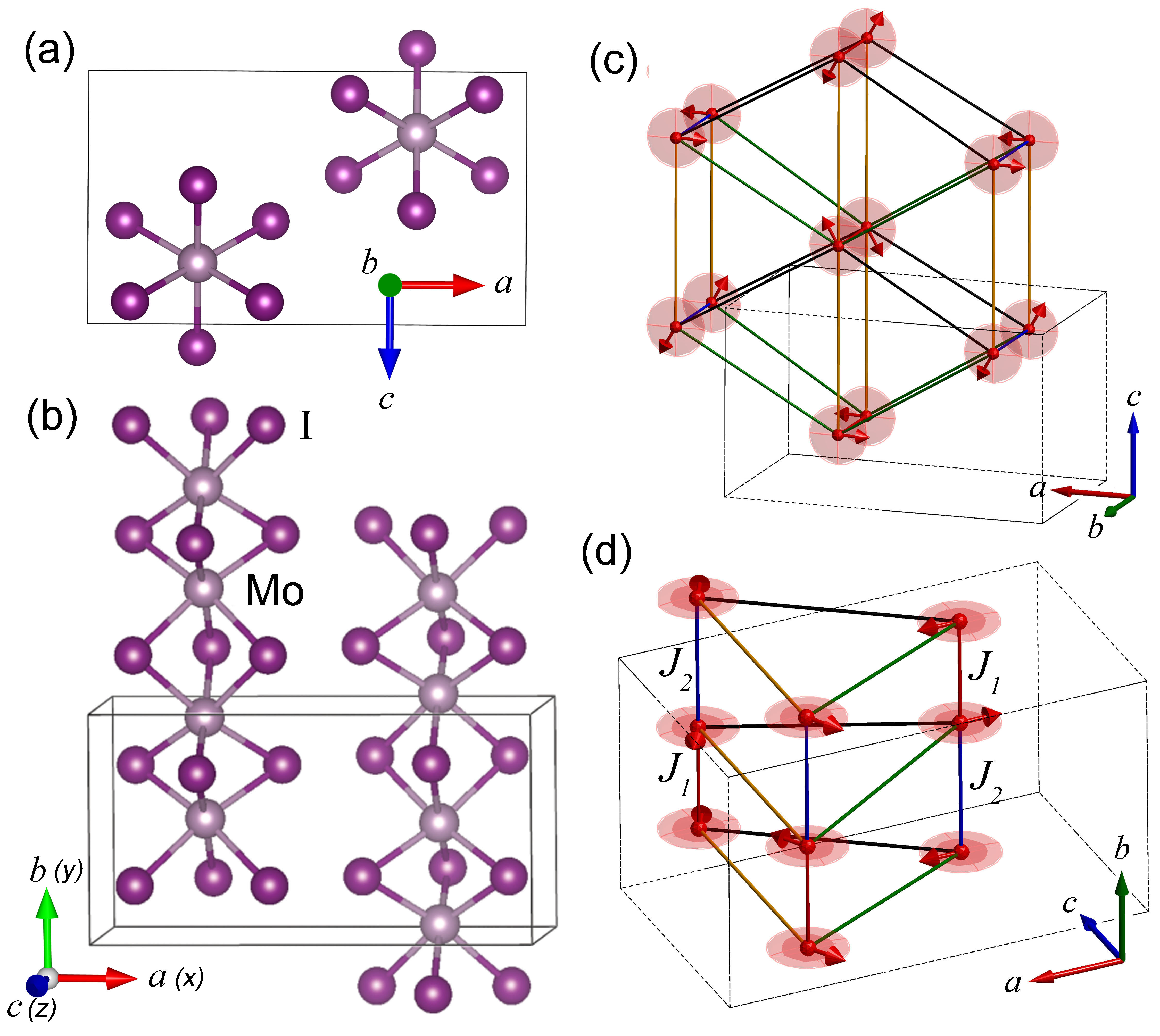}
\caption{
(a) Top view and (b) side view of two chains that comprise the MoI$_3$ unit cell. The chains have been doubled along
the $b$ axis to show the dimerization. 
The lattice vector directions, $xyz$ coordinate system, and outline of the unit cell are shown.
(c) Relaxed magnetic structure showing only the Mo atoms of a central chain and its six surrounding chains.
The circular regions around each spin indicate easy plane anisotropy perpendicular to the axis
of the chains.
(d) Relaxed magnetic structure illustrating the exchange coupling parameters.
Figures are generated by (a,b) Vesta \cite{VESTA} and (c,d) SpinW \cite{SpinW_Toth_2015}.
}
\label{fig:MoI3_struct_0}
\end{figure}

In this paper, we analyze the magnonic properties of a AFM dimerized and tetramerized spin chain.
Dimerization causes the AFM exchange coupling between the magnetic atoms along the chain to alternate in magnitude.
The two different intrachain exchange couplings, $J_1$ and $J_2$, are illustrated in Fig. \ref{fig:MoI3_struct_0}(d).
With $J_1 > J_2 > 0$ (AFM coupling) the magnonic spectrum remains trivial for all
positive values of $J_1$ and $J_2$.
If $J_2$ becomes negative, such that the sign of the exchange coupling along the chain
alternates,
then, depending on the relative magnitudes of $J_1$, $J_2$, and the anisotropy constants,
the magnonic spectrum can enter a non-trivial topological state that is qualitatively
different from that of the SSH model.

An example quasi-one-dimensional material that may exhibit these
magnonic topological  properties
is the quasi-one-dimensional transition metal tri-halide MoI$_3$.
It consists of covalently bonded MoI$_3$ chains weakly coupled to adjacent chains
in a triangular lattice. 
The material was recently synthesized and characterized with powder and single-crystal X-ray analysis
\cite{2016_MoI3_Facile_Synthesis,MoI3_exp,2022_MoI3_Fari_APL}. 
The crystal system is orthorhombic with space group {\em Pmmn} (59).
The chains are dimerized with alternating distances, 
2.88 {\AA} and 3.53 \AA, between Mo atoms along the chains.
There is no known temperature for the chains to transition to the symmetric phase
($P6_3/mmc$).
Each Mo atom is bonded to 6 I atoms in a distorted octahedral arrangement.
A bulk unit cell, doubled along the $b$ axis, is shown 
from top and side views in Fig. \ref{fig:MoI3_struct_0}(a,b).
Bulk samples were characterized by Raman spectroscopy \cite{2022_MoI3_Fari_APL}.
The temperature variation of several Raman peaks displayed a transition 
suggestive of a magnetic phase transition.
Several peaks that disappeared in the temperature range between 125 - 150 K were considered to originate
from two magnon scattering processes.
Density functional theory (DFT) calculations found that the lowest energy ground state 
was easy-plane AFM with alternating spins along the chains.
Prior DFT calculations of single chains of transition metal di- and tri-halides
also found the ground state of \MoI to be AFM 
\cite{2022_TM_Halide_Wires_APL,2024_MoI3_Spin_Chains_JMMM}.

Exchange coupling constants and anisotropy energies determined from DFT calculations
in Sec. \ref{sec:Methods_Models} 
are used in a Heisenberg type Hamiltonian that is analyzed using linear
spin wave theory \cite{Yosida_Th_Mag,Nolting_Qu_Mag,2020_Rezende_Fund_Magnonics}
in the main text and classical effective field theory in the Appendix.
Analytical expressions are derived in Sec. \ref{sec:Single_Chain} 
for an individual chain that explicitly show
the criteria for transitioning from a topologically trivial to a non-trivial phase.
The full magnonic spectrum that includes the interchain coupling is determined numerically
in Sec. \ref{sec:Interchain_Coupling}.

\section{Density Functional Theory Calculations}
\label{sec:Methods_Models}
Density functional theory calculations, 
as implemented in the Vienna $ab$ $initio$ simulation package (VASP) 
\cite{VASP:I:PRB:1993,VASP:II:PRB:1994,VASP:III:PRB:1996,VASP4_CompMatSci},
are performed using the generalized gradient approximation 
for the exchange-correlation functional as parameterized by Perdew, Burke, and Ernzerhof (PBE) \cite{PBE}.
The van der Waals interaction is included
using the PBE-D3 method of Grimme {\em et al.} \cite{grimme2010consistent}.
The energy cutoff for the plane wave basis is 520 eV.
Bulk structures are relaxed until the forces on the atoms are smaller than $10^{-4}$ eV/\AA.
Phonon calculations are performed using the finite-displacement supercell approach as implemented in 
Phonopy \cite{phonopy-phono3py-JPCM,phonopy-phono3py-JPSJ}
with a $2 \times 2 \times 2$ supercell,
and the Brillouin zone is sampled by a $3 \times 6 \times 6$ Monkhorst–Pack k-point grid.

\begin{figure}[t]
\centering
\includegraphics[width=2.8in]{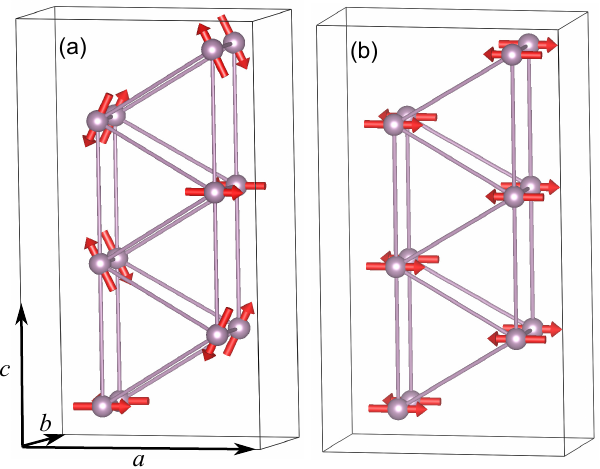}
\caption{
Magnetic configurations of the MoI$_3$ 1$\times$1$\times$3 supercell for the 
(a) non-collinear antiferromagnet state and 
(b) collinear antiferromagnet.
Only the Mo atoms are shown.}
\label{supercell113}
\end{figure}

\begin{figure}[t]
\centering
\includegraphics[width=3.3in]{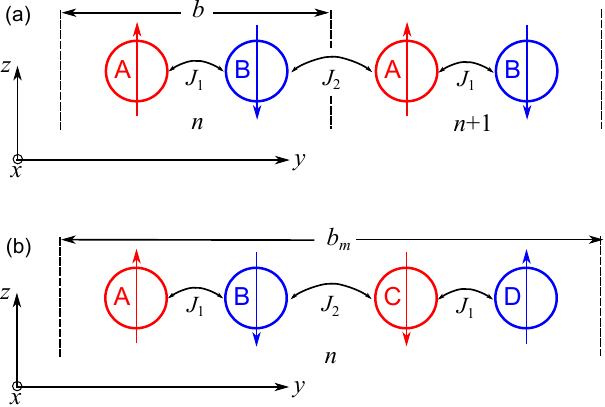}
\caption{
(a) Dimerized AFM chain with $J_1 , J_2 > 0$.
There are 2 Mo atoms per unit cell labeled as A and B. 
The unit cell indices shown are $n$ and $n+1$.
The lattice constant is $b$ for both the crystal lattice and the magnetic lattice.
The coordinate system corresponding to the crystallographic axes is also shown.
The chain lies along the $y$ direction.
The spins lie in the $x$-$z$ plane perpendicular to the chain axis.
We assume an arbitrarily small in-plane anisotropy and align the equilibrium spins along the $z$-axis.
(b) Magnetic unit cell with $J_1 > 0$ and $J_2 < 0$. The coupling now alternates between AFM and FM.
The magnetic unit cell length $b_m = 2b$ doubles so that there are 4 Mo atoms in the magnetic unit cell
labelled A - D. We refer to this as the magnetically tetramerized unit cell.
}
\label{1dimerized_2dimerized_chain}
\end{figure}

Previously, a Hubbard $U$ value of 4 eV was used to prevent the occurrence of negative phonon modes 
in the phonon dispersion calculations \cite{2022_MoI3_Fari_APL}.
In this work, we find that by using a tighter convergence criteria during the structure relaxation process,
a stable crystal structure can be obtained without including a Hubbard $U$ term.
The self consistent field electronic calculations with $U=0$, $U=2.4$ eV \cite{AFLOW_Hi_Thruput_CompMatSci_2015}, 
and $U=4$ eV \cite{2022_MoI3_Fari_APL}
all show the easy-plane AFM spin configuration, with alternating spins along the chains, 
to be the minimum energy magnetic ground state.
However, when the crystal structure is relaxed in the AFM phase, only the $U=0$ calculation results in the
dimerized {\em Pmmn} ground state.
The $U\neq 0$ calculations find the non-dimerized Pmmn phase to be the ground state, which contradicts the experimental results.
The total energy difference per unit cell of the dimerized and non-dimerized Pmmn phase (calculated with $U=0$ in the AFM phase) is 0.43 eV.
Furthermore, the $U=0$ calculation gives lattice constants (a = 12.41 \AA, b =6.45 \AA, and c = 7.17 \AA) that agree
closely with the experimental values \cite{2022_MoI3_Fari_APL}.
Therefore, all calculations in this work are performed without a Hubbard $U$ correction.
To investigate the inter-chain spin alignment of the charge neutral
ground state, 
we constructed a 1$\times$1$\times$3 supercell and compared the total energies 
of the non-collinear and collinear AFM states, as shown in Fig. \ref{supercell113}.
In the collinear configuration, the magnetic moments are aligned along the $\pm a$
axis.
In the non-collinear configuration, the magnetic moments of the Mo atoms in the same plane
that form the vertices of a triangle are rotated with respect to each other by
$120^\circ$ as shown in Fig. \ref{supercell113}(a). 
The DFT calculations reveal that the non-collinear AFM state is energetically favored by 
14.5 meV compared to the collinear AFM state.
This is also the minimum energy state found from the Heisenberg Hamiltonian (\ref{eq:dimerized_AFM_H})
using exchange parameters extracted from the DFT calculations.
The the magnetic exchange coupling constants
and anisotropy energies 
are determined from the standard energy mapping method 
\cite{2009_Whangbo_diamon_chain_JPCM,2022_Na2Cu2TeO6_Dagatto_PRB,2022_CrTe2_YLiu_PRMat,2022_MoI3_Fari_APL}
with total energies calculated using PBE-D3 with spin-orbit coupling (SOC). 
The details of the these calculations
and an illustration of the interchain exchange parameters, $J_3$ and $J_4$,
are provided in Appendix \ref{app:exchange}.
The results are shown in Table \ref{tab:parameters_filling}.
\begin{table}
\centering
\begin{tabular}{|l|ccccccc|}
\hline
Filling & Neutral & 0.1 & 0.19 & 0.196 & 0.2 & 0.3 & 0.4\\
\hline
$J_1$ & 24.08 & 23.80 & 23.86 & 23.88 & 23.89 & 24.39 & 25.42\\
\hline
$J_2$ & 4.44 & 2.04 & 0.02 & -0.11 & -0.20 & -2.30 & -4.26\\
\hline
$J_3$ & 0.55 & 0.61 & 0.67 & 0.68 & 0.68 & 0.74 & 0.82 \\
\hline
$J_4$ & 0.12 & 0.16 & 0.19 & 0.19 & 0.19 & 0.21 & 0.21 \\
\hline
$A_y$ & 2.11 & 2.38 & 2.57 & 2.58 & 2.58 & 2.76 & 2.90 \\
\hline
$A_x$ & 0 & 0.12 & 0.33 & 0.35 & 0.36 & 0.62 & 0.91 \\
\hline
\end{tabular}
\caption{Exchange and anisotropy parameters (meV) as a function of hole filling 
(number of excess holes per unit cell consisting of 4 Mo atoms and 12 I atoms).}
\label{tab:parameters_filling}
\end{table}
In the charge neutral state, all of the exchange coupling parameters are positive (AFM), 
and the minimum energy spin configuration along the chain, determined from DFT, is illustrated
in Fig. \ref{1dimerized_2dimerized_chain}(a).
With $A_y = 2.11$ meV and $A_x = 0$, the spins have 
easy-plane anisotropy with the easy-plane perpendicular to the chains, as illustrated
in Fig. \ref{fig:MoI3_struct_0}(c,d). 
With hole filling, $J_2$ changes sign and the minimum energy spin configuration along the chain switches to that
shown in Fig. \ref{1dimerized_2dimerized_chain}(b).
We will refer to this as the magnetically tetramerized unit cell.
Furthermore, the in-plane anisotropy $A_x$ increases, 
so that at larger values of hole filling,
the ground state spin configuration
becomes collinear with all spins aligned along $\pm \cv$.
We analyze the excitations of the magnetic system by first considering the individual chains
and then consider the effect of interchain coupling.

\section{Single Chain Analysis}
\label{sec:Single_Chain}
\subsection{Dimerized AFM Single Chain}
\label{sec:Dimerized_AFM_Struct}

We begin by analyzing a single chain illustrated in Fig. \ref{1dimerized_2dimerized_chain}(a)
using parameters for the charge neutral state given in Table \ref{tab:parameters_filling}.
We do this because the bulk crystal is physically a lattice of weakly coupled chains, 
and the characteristics of the single chain are still present and visible when interchain coupling is included.
Furthermore, the single chain provides a model system to compare against other such model systems found in the
literature, falling under the general umbrella of the SSH model.
The spin Hamiltonian for this system is
\be
\begin{aligned}
H &= 2 \sum_n \left[ J_1 \Sv_n^A \cdot \Sv_n^B + J_2 \Sv_n^A \cdot \Sv_{n-1}^B \right]\\
&+ A_y \sum_n \left[ \left(\Sv_{n,y}^A\right)^2  + \left(\Sv_{n,y}^B\right)^2 \right]\\
&+ A_x \sum_n \left[ \left(\Sv_{n,x}^A\right)^2  + \left(\Sv_{n,x}^B\right)^2 \right] .
\end{aligned}
\label{eq:dimerized_AFM_H}
\ee
In Eq. (\ref{eq:dimerized_AFM_H}), $J_1,J_2 > 0$, i.e. both exchange couplings are antiferromagnetic,
and $J_1 > J_2$.
$J_1$ is the exchange coupling between the $A$ and $B$ spins within the same unit cell,
and $J_2$ couples spins between adjacent unit cells.
The sum is over unit cells $n$.
$A_y \gg A_x > 0$ are anisotropy terms. 
$A_y$ creates easy-plane anisotropy with the easy plane in the $x$-$z$ plane perpendicular
the axis of the chain.
Within the $x$-$z$ plane a small anisotropy $A_x$ is assumed  
such that the equilibrium spins minimize their energy by aligning and anti-aligning along $z$.

The magnon excitation spectrum can be calculated using the classical effective field approach
or the quantum approach and linear spin wave theory \cite{Yosida_Th_Mag,Nolting_Qu_Mag,2020_Rezende_Fund_Magnonics}.
Here, we will use the quantum approach and linear spin wave theory, and, in Appendix 
\ref{app:Eff_Field_Magnon}, we describe the effective field approach.
Both approaches result in the same magnon dispersion, however the quantum approach
allows for the calculation of the surface Green's function using the standard
decimation algorithm \cite{Sancho_Rubio_JPhysF84,MPLSancho_HighlyConvergent_JPF85,Galperin:JCP:2002:decimation}.

\begin{figure}[tb]
\centering
\includegraphics[width=3.3in]{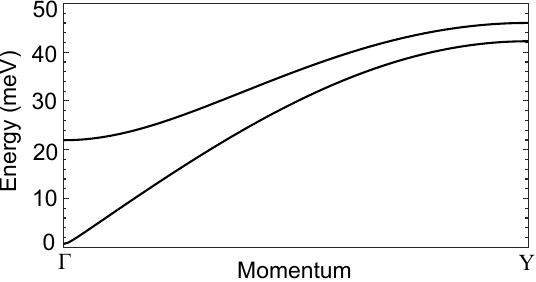}
\caption{
Magnon spectrum of a single 1D dimerized AFM chain calculated from Eqs. (\ref{eq:w0-k_1D_dimerized_AFM})
and (\ref{eq:w1-k_1D_dimerized_AFM}).
The values of $J_1$, $J_2$, and $A_y$ are from the `Neutral' column of Table
\ref{tab:parameters_filling}.
In addition, $A_x$ is set to 2 $\mu$eV to show the effect of small in-plane
anisotropy.
}
\label{fig:Filling5_neutral_w-k_chain_bulk}
\end{figure}

The analysis begins with the Holstein-Primakoff transformation \cite{Holstein_Primakoff1940}
of the spin operators in Eq. (\ref{eq:dimerized_AFM_H}).
To lowest order in the spin deviation operators,
\begin{align}
\hat{S}_{j,z}^{A} & = S - \nh^A_j \nonumber \\
\hat{S}_j^{A+} & \approx \sqrt{2S} a_j \nonumber \\
\hat{S}_j^{A-} & \approx \sqrt{2S} \; \ajd 
\label{eq:HPA_1st_order}
\end{align}
and
\begin{align}
\hat{S}_{j,z}^{B} & = -S + \nh^B_j \nonumber \\
\hat{S}_j^{B+} & \approx \sqrt{2S} \; b_j^\dagger \nonumber \\
\hat{S}_j^{B-} & \approx \sqrt{2S} b_j 
\label{eq:HPB_1st_order}
\end{align}
where $\nh^A_j = \ajd a_j$, $\nh^B_j = b_j^\dagger b_j$ and 
$n_j \in \{0, 1, 2, \ldots , 2S\}$ so that $S^{A}_{j,z}$ and $S^{B}_{j,z}$ run from $-S$ to $S$.
The creation and annihilation operators follow the usual bosonic commutation relations.
Making the above substitutions into Eq. (\ref{eq:dimerized_AFM_H}), 
using $S_x = \frac{1}{2} \left( S^+ + S^- \right)$ and
$S_y = \frac{1}{2i} \left( S^+ - S^- \right)$, and keeping terms up to quadratic order, we have,
$H \equiv H_{J_1} + H_{J_2} + H_A$ where
\begin{widetext}
\begin{align}
H_{J_1}&= 2 J_1 S \left\{ -NS + \sum_n \left( a_n b_n + a_n^\dagger b_n^\dagger + a_n^\dagger a_n + b_n^\dagger b_n \right)
\right\} \label{eq:HJ1}\\
H_{J_2}&= 2 J_2 S \left\{ -NS + \sum_n \left( a_n b_{n-1} + a_n^\dagger b_{n-1}^\dagger + a_n^\dagger a_n + b_{n-1}^\dagger b_{n-1} \right) \right\} \label{eq:HJ2}\\
H_A &= -\tfrac{A_y - A_x}{2} S \sum_n \left( a_n a_n + a_n^\dagger a_n^\dagger + b_n b_n + b_n^\dagger b_n^\dagger \right)
 + \tfrac{A_y + A_x}{2} \sum_n \left( a_n a_n^\dagger + a_n^\dagger a_n + b_n^\dagger b_n + b_n b_n^\dagger \right) .
\label{eq:HA}
\end{align}
Defining the Fourier transforms as
$a_n = \frac{1}{\sqrt{N}} \sum_{q} e^{i q b n} a_q$, 
and
$b_n = \frac{1}{\sqrt{N}} \sum_{q} e^{i q b n} b_q$,
where $b$ is the lattice constant of the 
1D chain shown in Fig. \ref{1dimerized_2dimerized_chain}(a),
$n$ is the unit cell index, 
and $q \in \{-\pi/b, \pi/b\}$,
the transformed Eqs. (\ref{eq:HJ1})-(\ref{eq:HA}) become
\begin{equation}
\begin{aligned}
H_{J_1} =& 2J_1S \left\{ -NS
+ \sum_q \left(
a_{-q} b_{q} + a_q^\dagger b_{-q}^\dagger + a_q^\dagger a_q + b_q^\dagger b_q 
\right) \right\}\\
H_{J_2} =& 
2J_2S \left\{
-NS + \sum_q \left(
e^{iqb} a_q b_{-q} + e^{-iqb} a_q^\dagger b_{-q}^\dagger + a_q^\dagger a_{q} + b_q^\dagger b_{q}
\right)
\right\}\\
H_A =& -\tfrac{A_y-A_x}{2}S \sum_q \left( a_q a_{-q} + a_q^\dagger a_{-q}^\dagger + b_q b_{-q} + b_q^\dagger b_{-q}^\dagger 
\right)
+ \tfrac{A_y + A_x}{2} S \sum_q \left( a_q a_q^\dagger + a_q^\dagger a_q + b_q^\dagger b_q + b_q b_q^\dagger \right) .
\end{aligned}
\label{H2nd(q)_dimerized_AFM_Ax_Ay}
\end{equation}
There is considerable flexibility in the way we write Eq. (\ref{H2nd(q)_dimerized_AFM_Ax_Ay}), since for any term, we can let
$q \rightarrow -q$, and we can also apply the commutation relations to re-arrange any pair of operators.
We use this flexibility 
to arrange the Hamiltonian into a manifestly Hermitian form.
\begin{equation}
\begin{aligned}
H & = - 4S \left( J_1 + J_2 \right) NS \\
&+ S \sum_q \left[ \left( J_1 + J_2 e^{iqb} \right) \left( b_{-q}a_q + b_q^\dagger a_{-q}^\dagger \right)  
+ \left( J_1 + J_2 e^{-iqb} \right) \left( a_q^\dagger b_{-q}^\dagger  + a_{-q}{b_q} \right) \right]\\
&+ S \sum_q \left[ \left(J_1 + J_2 + \tfrac{A_y + A_x}{2} \right) 
\left( a_q^\dagger a_q + a_{-q} a_{-q}^\dagger + b_q^\dagger b_q + b_{-q}b_{-q}^\dagger \right) \right]
- \tfrac{A_y-A_x}{2}S \sum_q \left( a_{-q} a_{q} + a_q^\dagger a_{-q}^\dagger + b_{-q} b_{q} + b_q^\dagger b_{-q}^\dagger 
\right)
\end{aligned}
\label{eq:H2nd(q)_dimerized_AFM_Ax_Ay_manifest_H_factored_w_H22}
\end{equation}

We now follow an approach described by
White {\em et al}. \cite{1965_Diag_AFM_Magnon_Phonon_Interaction}
and Colpa \cite{1978_Colpa},
which is a common approach for calculating magnon energies in AFMs
\cite{SpinW_Toth_2015,2020_Rezende_Fund_Magnonics,2022_AFM_Nernst_Cheng_APL}.
Ignoring the constant term, we write $H$ as
\begin{equation}
H = \tfrac{1}{2}\sum_q \Xv^\dagger(q) \Hv(q) \Xv(q)
\label{eq:XHX}
\end{equation}
where $\Xv$ is a column vector of 4 operators, and $\Hv$ is a $4 \times 4$ matrix of $c$-numbers.
We choose
$\Xv = [ a_q \;\; b_q \;\; a_{-q}^\dagger \;\; b_{-q}^\dagger ]^T$.
$\Xv$ satisfies the matrix commutation relation
$\left[ \Xv, \Xv^\dagger \right] \equiv \Xv \left( \Xv^* \right)^T - \left( \Xv^* \Xv^T \right)^T \equiv \gv_a
= 
\sigma_z \otimes \Iv$
where $\sigma_z$ is the Pauli matrix, $\Iv$ is the $2\times 2$ identity matrix,
$\Xv^*$ is the column vector $\Xv$ with the operators daggered, 
and $(\Xv^*)^T = \Xv^\dagger$. 
The matrix $\Hv(q)$ is
\begin{equation}
\Hv(q) = 
2S \left[
\begin{array}{cccc}
\Sg & 0 & - \Delta_A &  \left(J_1 + J_2 e^{-iqb} \right)\\
0 & \Sg &  \left(J_1 + J_2 e^{iqb} \right) & - \Delta_A\\
- \Delta_A &  \left(J_1 + J_2 e^{-iqb} \right) & \Sg & 0\\
 \left(J_1 + J_2 e^{iqb}\right) & - \Delta_A & 0 & \Sg
\end{array}
\right]
\label{eq:Hq_dimerized_AFM_w_A_and_H22}
\end{equation}
where $\Sg = \left(J_1 + |J_2| + \tfrac{A_y + A_x}{2} \right)$ and $\Delta_A =  \tfrac{A_y-A_x}{2}$.
The eigenenergies of the excitations are determined from the generalized eigenvalue equation
$\left| \Hv(q) - \gv_a \hbar \w \right| = 0$.
Since $\gv_a$ is invertible, and $\gv_a^{-1} = \gv_a$, this is also easily converted into a regular eigenvalue equation,
$\left| \Hv(q) \gv_a -  \Iv \hbar \w  \right| = 0$
where $\Iv$ is the identity matrix.
There are 4 solutions of the form $\pm \hbar \w_0$ and $\pm \hbar \w_1$. 
The two positive solutions give the $\w-k$ relations of the two magnon modes.
The dispersion relations for the two positive energy modes are
\be
\hbar \w_0(k) = 
2S 
\left[ 
4 J_1 J_2 \sin^2(\tfrac{kb}{2}) 
- (A_y-A_x) (J_1+J_2) \sqrt{1 - \frac{4J_1J_2\sin^2(\tfrac{kb}{2})}{(J_1+J_2)^2}}
+ (A_y+A_x) (J_1 + J_2) 
+ A_yA_x
\right]^{1/2} \; 
\label{eq:w0-k_1D_dimerized_AFM}
\ee
and
\be
\hbar w_1(k) = 
2S \left[ 4 J_1 J_2 \sin^2(\tfrac{kb}{2}) + (A_y-A_x)(J_1+J_2) 
\sqrt{1 - \frac{4J_1J_2\sin^2(\tfrac{kb}{2})}{(J_1+J_2)^2}}
+  (A_y+A_x) (J_1 + J_2) + A_yA_x
\right]^{1/2} \; .
\label{eq:w1-k_1D_dimerized_AFM}
\ee
\end{widetext}
The dispersion of the single dimerized AFM chain, 
calculated from Eqs. (\ref{eq:w0-k_1D_dimerized_AFM}) and (\ref{eq:w1-k_1D_dimerized_AFM}), 
is plotted in Fig. \ref{fig:Filling5_neutral_w-k_chain_bulk}(a) 
using parameters from Table \ref{tab:parameters_filling} for the charge neutral structure.
To show the effect of in-plane anisotropy, a value of $A_x = 0.002 \; {\rm meV} \: \approx A_y/1000$ is also used.
In the absence of anisotropy ($A_y = A_x = 0$), the dispersions of both modes 
reduce to the standard dispersion of a 
1D AFM chain \cite{Kittel} with the uniform exchange coupling replaced by the geometric mean 
of the two different exchange couplings,
\be
\hbar \w_0(k) = 4S\sqrt{J_1 J_2} \sin\left(\tfrac{kb}{2}\right). 
\ee
Unlike the ferromagnetic chain, 
in which dimerization gaps the magnon spectrum in the middle of the band in analogy with 
the Su-Schrieffer-Heeger (SSH) model for electrons \cite{2018_SSH_Magnon_Chain}, 
the dimerization of an AFM chain does not qualitatively alter the spectrum.
It remains linear and gapless.

The energies at $\Gamma$ are
\be
\hbar \w_0 = 2S \sqrt{2 A_x (J_1 + J_2 + A_y/2)} \approx 2S \sqrt{2A_x(J_1+J_2)}
\ee
and
\be
\hbar \w_1 = 2S \sqrt{2A_y (J_1 + J_2 + A_x/2)} \approx 2S \sqrt{2A_y(J_1+J_2)} \; .
\ee
Thus, even though the in-plane anisotropy $A_x$ may be extremely small,
the fundamental gap $\hbar \w_0$ in the spectrum is determined by the geometric mean of the in-plane anisotropy
and the exchange coupling, and this can be 2 orders of magnitude larger than the small anisotropy energy.

At the zone boundary ($k = \pi/b$), the energies are
\be
\hbar \w_0(\pi/b) = 4S\sqrt{J_1J_2}
\left[ 1 + \frac{A_yJ_2 + A_xJ_1 + A_yA_x/2}{2J_1J_2} \right]^{1/2}
\ee
\be
\hbar \w_1(\pi/b) = 4S\sqrt{J_1J_2} \left[ 1 + \frac{A_yJ_1 + A_xJ_2 + A_yA_x/2}{2J_1J_2} \right]^{1/2}
\ee
Expanding out the square root to first order,
\be
\hbar \w_0(\pi/b) \approx 4S\sqrt{J_1J_2} \left\{
1 + \frac{A_yJ_2 + A_xJ_1 + A_yA_x/2}{4J_1J_2} \right\}
\ee
\be
\hbar \w_1(\pi/b) \approx 4S\sqrt{J_1J_2} \left\{
1 + \frac{A_yJ_1 + A_xJ_2 + A_yA_x/2}{4J_1J_2} \right\} ,
\ee
and taking the difference of the above two equations gives
\be
\hbar \w_1(\pi/b) - \hbar \w_0(\pi/b) = 
\frac{S(J_1-J_2)(A_y-A_x)}{\sqrt{J_1J_2}} .
\label{eq:zone_edge_splitting}
\ee
From Eq. (\ref{eq:zone_edge_splitting}), 
we see that splitting of the two modes at the zone edge requires both dimerization 
($J_1 \neq J_2$) and anisotropy.
\subsection{Tetramerized AFM-FM Single Chain}
\label{sec:AFM-FM_Single_Chain}
When $J_2$ changes sign, the magnetic ground state of the single chain becomes that shown in
Fig. \ref{1dimerized_2dimerized_chain}(b).
The spin Hamiltonian retains the same form as Eq. (\ref{eq:dimerized_AFM_H}), except now, $J_1 > 0$, $J_2 < 0$,
and there are 4 atoms in the magnetic unit cell.
The Hamiltonian is now
\small
\be
\begin{aligned}
H   = & 2  \sum_n 
\left[ 
J_1 \Sv_n^A \cdot \Sv_n^B + 
J_2 \Sv_n^B \cdot \Sv_{n}^C 
+   J_1 \Sv_n^C \cdot \Sv_n^D + 
J_2 \Sv_n^D \cdot \Sv_{n+1}^A \right] \\
&
+ A_y \sum_{n, \alpha} \left( \Sv_{n,y}^\alpha \right)^2  
+ A_x \sum_{n, \alpha} \left(\Sv_{n,x}^\alpha \right)^2  ,
\end{aligned}
\label{eq:doubly_dimerized_AFM_H}
\ee
\normalsize
where $\alpha$ is the atom index $\{A, B, C, D\}$.
The Holstein-Primakoff transformations for the spin operators $\Sv^A$ and $\Sv^B$ remain the same as
given by Eqs. (\ref{eq:HPA_1st_order}) and (\ref{eq:HPB_1st_order}).
The Holstein-Primakoff transformations for $\Sv^C$ and $\Sv^D$ are
\begin{equation}
\begin{aligned}
\hat{S}_{j,z}^{C} & = -S + \nh^C_j; \;\;
\hat{S}_j^{C+}  \approx \sqrt{2S} \; c_j^\dagger; \;\;
\hat{S}_j^{C-}  \approx \sqrt{2S} c_j \\ 
\\
\hat{S}_{j,z}^{D} & = S - \nh^D_j; \;\;
\hat{S}_j^{D+}  \approx \sqrt{2S} d_j; \;\;
\hat{S}_j^{D-}  \approx \sqrt{2S} \; d_j^\dagger
\label{eq:HPABCD_1st_order}
\end{aligned}
\end{equation}
\begin{widetext}
Writing out $H$ in Eq. (\ref{eq:doubly_dimerized_AFM_H}) as the parts proportional to 
$J_1$, $J_2$, and the anisotropy terms gives
\begin{equation}
\begin{aligned}
H_{J_1} &= 2 J_1 S 
\left\{ 
-2NS + \sum_n \left( 
a_n b_n + a_n^\dagger b_n^\dagger + a_n^\dagger a_n + b_n^\dagger b_n +
c_n^\dagger d_n^\dagger + c_n d_n + c_n^\dagger c_n + d_n^\dagger d_n
\right)
\right\} 
\\
H_{J_2} &= 2 |J_2| S \left\{ -2NS + \sum_n 
\left( -b_n^\dagger c_{n} - b_n c_{n}^\dagger + b_n^\dagger b_n + c_{n}^\dagger c_{n} -
d_n a^\dagger_{n+1} - d_n^\dagger a_{n+1} + a^\dagger_{n+1} a_{n+1} +  d^\dagger_{n} d_{n}
\right) 
\right\}
\\
H_A &= -\tfrac{A_y - A_x}{2} S \sum_n 
\left( 
a_n a_n + a_n^\dagger a_n^\dagger + b_n b_n + b_n^\dagger b_n^\dagger + c_n c_n + c_n^\dagger c_n^\dagger + d_n d_n + d_n^\dagger d_n^\dagger 
\right) 
\\
&+
\tfrac{A_y + A_x}{2} \sum_n 
\left( 
a_n a_n^\dagger + a_n^\dagger a_n + b_n^\dagger b_n + b_n b_n^\dagger + c_n c_n^\dagger + c_n^\dagger c_n + d_n^\dagger d_n + d_n d_n^\dagger 
\right) .
\end{aligned}
\label{eq:HA_DD}
\end{equation}
Fourier transforming, we have
\begin{equation}
\begin{aligned}
H =& 2J_1S \left\{ -2NS
+ \sum_q \left(
a_{q} b_{-q} + a_q^\dagger b_{-q}^\dagger + 
c_q d_{-q} + c_q^\dagger d_{-q}^\dagger +
a_q^\dagger a_q + b_q^\dagger b_q  + 
c_q^\dagger c_q + d_q^\dagger d_q  \right) \right\}
\\
+& 2 |J_2| S \left\{ -2NS + \sum_q \left(
-b_q^\dagger c_q - b_q c_{q}^\dagger  
- e^{-iqa} d_q a_{q}^\dagger - e^{iqa} d_q^\dagger a_{q} + a_q^\dagger a_{q} + b_q^\dagger b_{q} + c_q^\dagger c_{q} + d_q^\dagger d_{q}
\right) \right\}
\\
-& 
\tfrac{A_y-A_x}{2}S \sum_q \left( 
a_q a_{-q} + a_q^\dagger a_{-q}^\dagger + b_q b_{-q} + b_q^\dagger b_{-q}^\dagger + 
c_q c_{-q} + c_q^\dagger c_{-q}^\dagger + d_q d_{-q} + d_q^\dagger d_{-q}^\dagger 
\right)
\\
&+ \tfrac{A_y + A_x}{2} S \sum_q 
\left( a_q a_q^\dagger + a_q^\dagger a_q + b_q^\dagger b_q + b_q b_q^\dagger + c_q c_q^\dagger + c_q^\dagger c_q + d_q^\dagger d_q + d_q d_q^\dagger 
\right) .
\end{aligned}
\label{eq:H2nd(q)_DD}
\end{equation}
We again symmetrize terms to make the Hamiltonian manifestly Hermitian.
\begin{equation}
\begin{aligned}
H =& - 4 N S^2 \left( J_1 + |J_2| \right)  \\
+& S \left(J_1 + |J_2| + \tfrac{A_y + A_x}{2} \right) 
\sum_q \left( 
a_q^\dagger a_q + a_{-q}  a_{-q}^\dagger +
b_q^\dagger b_q  + b_{-q} b_{-q}^\dagger +
c_q^\dagger c_q + c_{-q}  c_{-q}^\dagger +
d_q^\dagger d_q  + d_{-q}  d_{-q}^\dagger \right)
\\
+&  J_1 S \sum_q \left( 
a_{-q} b_{q} + b_{-q} a_{q} +
a_q^\dagger b_{-q}^\dagger + b_{q}^\dagger a_{-q}^\dagger + 
c_{-q} d_{q} + d_{-q} c_{q} + 
c_q^\dagger d_{-q}^\dagger + d_{q}^\dagger c_{-q}^\dagger 
\right) 
\\
+& |J_2| S \sum_q \left(
-b_q^\dagger c_q - c_{-q} b_{-q}^\dagger 
- b_{-q} c_{-q}^\dagger  - c_{q}^\dagger   b_{q} 
- e^{iqa} d_{-q} a_{-q}^\dagger - e^{-iqa} a_{q}^\dagger d_{q} 
- e^{iqa} d_q^\dagger a_{q} - e^{-iqa} a_{-q} d_{-q}^\dagger 
\right) 
\\
+& 
-\tfrac{A_y-A_x}{2}S \sum_q \left( 
a_{-q} a_{q} + a_q^\dagger a_{-q}^\dagger + b_{-q} b_{q} + b_q^\dagger b_{-q}^\dagger + 
c_{-q} c_{q} + c_q^\dagger c_{-q}^\dagger + d_{-q} d_{q} + d_q^\dagger d_{-q}^\dagger 
\right)
\end{aligned}
\label{eq:H2nd(q)_DD2}
\end{equation}
\end{widetext}

\begin{figure}[thb]
\centering
\includegraphics[width=3.3in]{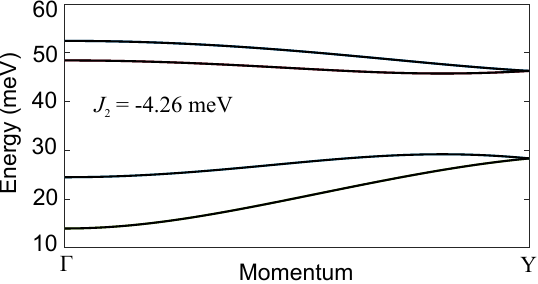}
\caption{
Magnon spectrum for the magnetic tetramerized 1D chain shown in 
Fig. \ref{1dimerized_2dimerized_chain}(b) calculated from Eqs. (\ref{eq:E-k_1,2_dgafm}) 
and (\ref{eq:E-k_3,4_dgafm})
with parameters from Table \ref{tab:parameters_filling} for 0.4 hole filling:
$J_1 = 25.42$ meV, $J_2 = -4.26$ meV, $A_y = 2.9$ meV, and $A_x = 0.91$ meV.
Y corresponds to the Brillouin zone edge of the magnetic unit cell,
${\rm Y} = \frac{\pi}{b_m} = \frac{\pi}{2b}$, where $b_m$ is the magnetic unit cell lattice constant shown 
in Fig. \ref{1dimerized_2dimerized_chain}(b).
}
\label{fig:E-k_doubly_dimerized_afm}
\end{figure}

We now write $H$ in the form of Eq. (\ref{eq:XHX}) with 
$\Xv^\prime = [a_q \; a_{-q}^\dagger \; b_q \; b_{-q}^\dagger \; c_q \; c_{-q}^\dagger \; d_q \; d_{-q}^\dagger ]^T$.
We choose this ordering in preparation for the calculation of the surface Green's function,
since it places the inter-unit-cell coupling at the upper left and lower right diagonal blocks in
the lattice representation.
$\Xv^\prime$ satisfies commutation relation $\left[ \Xv^\prime, {\Xv^\prime}^\dagger \right] = \gv_b$,
where $\gv_b = \Iv \otimes \sigma_z$, 
$\Iv$ is the $4\times 4$ identity matrix, and $\sigma_z$ is the Pauli matrix.
Hamiltonian (\ref{eq:H2nd(q)_DD2}) results in the following matrix $\Hv(q)$,
\begin{widetext}
\begin{equation}
\Hv(q) =
2S \left[
\begin{array}{cccccccc}
\Sg  & -\Delta_A  & 0 & J_1 & 0 & 0 & -|J_2|e^{-iqb_m} & 0 \\
-\Delta_A & \Sg  & J_1  & 0 & 0 & 0  & 0 & -|J_2|e^{-iqb_m} \\
0 & J_1 & \Sg  & -\Delta_A & -|J_2| & 0 & 0 &  0  \\
J_1 & 0 & - \Delta_A & \Sg  & 0 & -|J_2| & 0 &  0 \\
0 & 0 & -|J_2| & 0 & \Sg & -\Delta_A & 0 & J_1 \\
0 & 0 & 0 & -|J_2| & -\Delta_A & \Sg & J_1 & 0 \\ 
-|J_2|e^{+iqb_m} & 0 & 0 & 0 & 0 & J_1 & \Sg & -\Delta_A \\ 
0 & -|J_2|e^{+iqb_m} & 0 & 0 & J_1 & 0 & -\Delta_A & \Sg
\end{array}
\right]
\label{eq:Hp}
\end{equation} 
The eigenenergies, found from 
$\left| \Hv(q) - \gv_b \hbar \w \right| = 0$  
or
$\left| \Hv(q) \gv_b - \Iv \hbar \w \right| = 0$, 
again come in pairs $\pm \hbar \w(k)$.
The two lower positive bands are given by
\begin{equation}
\begin{aligned}
\hbar \w_{1,2}(k) = & 2S \Biggl\{ \; (A_x + A_y - 2J_2)(J_1 - J_2) + A_xA_y 
\\
& -
\Bigl[ \; 4J_1 J_2^2 (A_x + A_y - 2J_2) +  J_2^2 (A_x + A_y - 2J_2)^2 J_1^2 (A_y-A_x)^2 + 2 J_1^2 J_2^2(1+\cos(kb_m))
\\
& \pm
2 \left( \; \tfrac{1}{2} J_1^2 J_2^2 \left(A_x-A_y\right)^2 \left(A_x+A_y+2J_1-2J_2\right)^2(1+\cos(kb_m) \; \right)^{1/2}
\; \Bigr]^{1/2}
\; \Biggr\}^{1/2}
\end{aligned}
\label{eq:E-k_1,2_dgafm}
\end{equation}
where the plus sign corresponds to band 1 and the minus sign to band 2.
The two upper positive bands are given by
\begin{equation}
\begin{aligned}
\hbar \w_{3,4}(k) = & 2S \Biggl\{ \; (A_x + A_y - 2J_2)(J_1 - J_2) + A_xA_y 
\\
& +
\Bigl[ \; 4J_1 J_2^2 (A_x + A_y - 2J_2) +  J_2^2 (A_x + A_y - 2J_2)^2 J_1^2 (A_y-A_x)^2 + 2 J_1^2 J_2^2(1+\cos(kb_m))
\\
& \mp
2 \left( \; \tfrac{1}{2} J_1^2 J_2^2 \left(A_x-A_y\right)^2 \left(A_x+A_y+2J_1-2J_2\right)^2(1+\cos(kb_m) \; \right)^{1/2}
\; \Bigr]^{1/2}
\; \Biggr\}^{1/2}
\end{aligned}
\label{eq:E-k_3,4_dgafm}
\end{equation}
\end{widetext}
where the minus sign corresponds to band 3 and the plus sign to band 4.
Using parameters corresponding to a hole doping of 0.4 per bulk unit cell (4 Mo atoms and 12 I atoms),
$J_1 = 25.42$ meV, $J_2 = -4.26$ meV, $A_y = 2.9$ meV, and $A_x = 0.91$ meV, 
the positive valued bands are shown in Fig. \ref{fig:E-k_doubly_dimerized_afm}.

\begin{figure}[thb]
\centering
\includegraphics[width=3.3in]{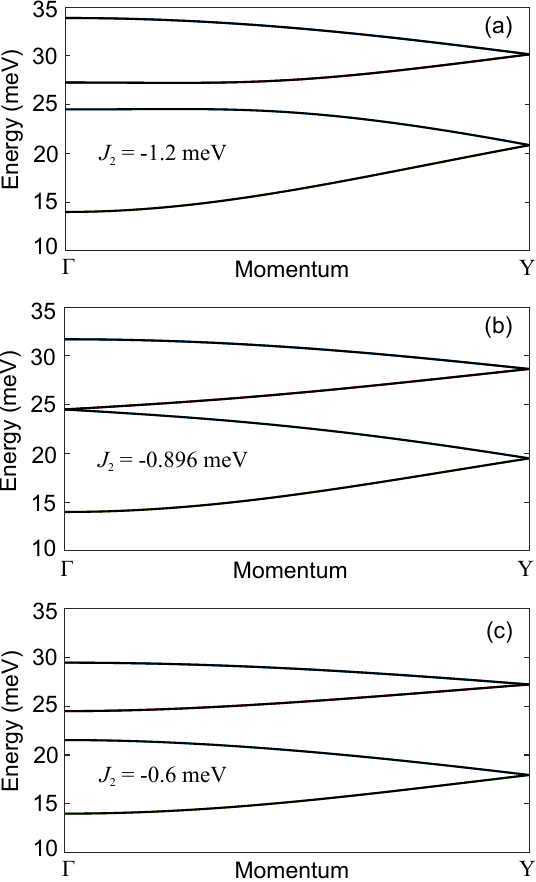}
\caption{
Magnon spectra as $J_2$ is changed from (a) $-1.2$ meV to (b) $-0.896$ meV to (c) $-0.6$ meV. 
The values for $J_1$, $A_y$, and $A_x$ remain the same as in Fig. \ref{fig:E-k_doubly_dimerized_afm}.
}
\label{fig:2_dimer_E-ks_3}
\end{figure}

The spectrum is now gapped, and the question is whether this gap is a trivial gap or a topological gap.
The first indication that there is a topologically non-trivial phase is that when we continually reduce the magnitude of
$J_2$, the gap in the spectrum closes and re-opens at $\Gamma$ as shown in Fig. \ref{fig:2_dimer_E-ks_3}. 
From the analytical expressions (\ref{eq:E-k_1,2_dgafm}) and (\ref{eq:E-k_3,4_dgafm}) evaluated at $k=0$, 
the energies of the 4 bands at $\Gamma$ are
\begin{equation}
\begin{aligned}
E_1 &= 2S \sqrt{A_x (A_y + 2J_1)} \\
E_2 &= 2S \sqrt{A_y (A_x + 2J_1)} \\
E_3 &= 2S \sqrt{(A_x + 2|J_2|)[A_y + 2(J_1 + |J_2|)]} \\
E_4 &= 2S \sqrt{(A_y + 2|J_2|)[A_x + 2(J_1 + |J_2|)]}. 
\end{aligned}
\label{eq:Ei_at_Gamma}
\end{equation}
Note that $E_2$ is independent of $J_2$. 
As $|J_2|$ is reduced, $E_3 = E_2$ when $J_2$ reaches a critical value $J_{2c}$ given by
a quadratic equation, which, when expanded out to first order in the small parameter
$\frac{(Ay-Ax)}{J_1\left(1 + (A_y + A_x)/(2J_1)\right)^2}$, gives
\begin{equation}
J_{2c} \approx - J_1 \frac{A_y - A_x}{2J_1 + A_x + A_y} \approx
-\frac{A_y - A_x}{2} + \frac{A_y^2 - A_x^2}{4 J_1} \: .
\label{eq:Jcrit}
\end{equation}
It is clear from Eq. (\ref{eq:Jcrit}) that both anisotropy and dimerization 
are required for a band crossing to occur.
For the values of $J_1$, $A_y$, and $A_x$ listed in Fig. \ref{fig:E-k_doubly_dimerized_afm},
corresponding to $-0.4$ filling in Table \ref{tab:parameters_filling}, the
exact value for $J_{2c}$ is $-0.896$ meV and the value for $J_{2c}$ from the linearized
expression is $-0.926$ meV.
The question is now, what is the nature of the bands on either side of the critical value $J_{2c}$?
On which side is the gap topological and on which side is it trivial?

\begin{figure}[thb]
\centering
\includegraphics[width=3.2in]{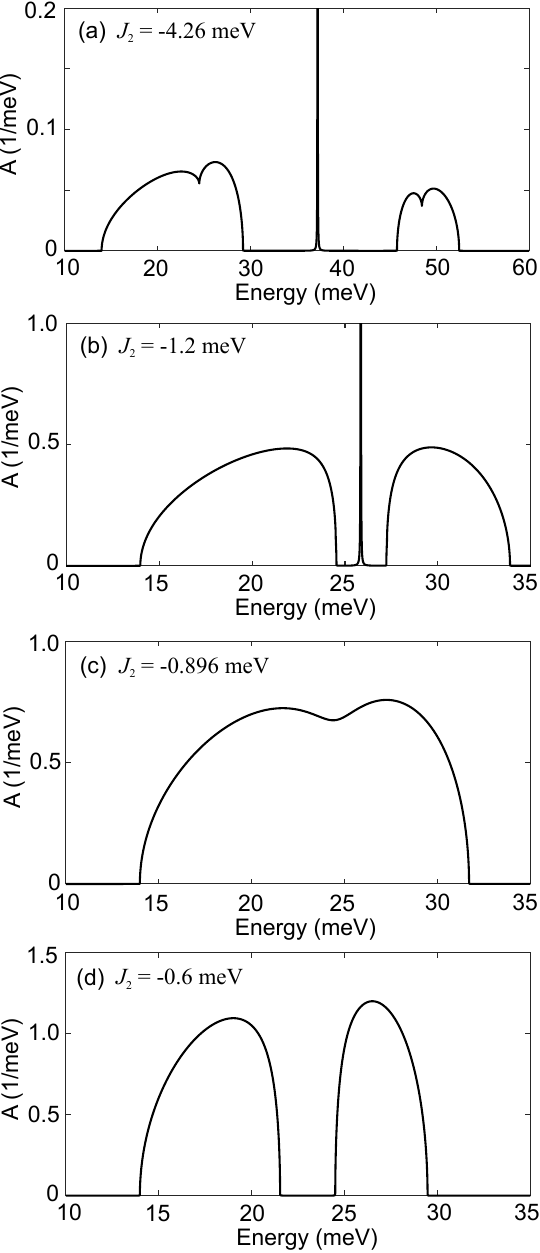}
\caption{
(a-d) Surface spectral functions 
calculated for the values of $J_2$ shown in the plots.
These values correspond to  those used  in the
$E-k$ plots shown in Figs. \ref{fig:E-k_doubly_dimerized_afm} and \ref{fig:2_dimer_E-ks_3}.
The values for $J_1$, $A_y$, and $A_x$ remain the same as in Fig. \ref{fig:E-k_doubly_dimerized_afm}.
In (a) and (b), the $y$-axis has been truncated, so that the band contributions to $A$ are visible.
The height of the peaks of the surface states are on the order of $10^3$ (1/meV).
}
\label{fig:2_dimer_As-As_3}
\end{figure}

To answer this question, we consider the `surface Green's functions'
and the corresponding `surface spectral functions' for
the 4 values of $J_2$ used in Figs. \ref{fig:E-k_doubly_dimerized_afm} and \ref{fig:2_dimer_E-ks_3}.
To obtain the surface Green function, we inverse Fourier transform $\Hv^\prime \equiv \Hv(q) \gv_b$ back
to the lattice representation 
\begin{equation}
\tv^\dagger \Uv_{n-1} + (\Dv-\tfrac{E}{2}\Iv) \Uv_n + \tv \Uv_{n+1} = 0
\label{eq:2_degen_H_symbolic}
\end{equation}
where $\Uv$ is now a column of the transformation matrix that diagonalizes 
$\Hv^\prime$ \cite{1965_Diag_AFM_Magnon_Phonon_Interaction}.
The matrices $\Dv$ and $\tv$ are
\be
\Dv = 
\left[
\begin{array}{cccccccc} 
\Sg            & \Delta_A      & 0          & -J_1     & 0         & 0        & 0              & 0 \\
-\Delta_A      & -\Sg          & J_1        & 0        & 0         & 0        & 0              & 0 \\
0              & -J_1          & \Sg        & \Delta_A & -|J_2|    & 0        & 0              &  0  \\
J_1            & 0             & - \Delta_A & -\Sg     & 0         & |J_2|    & 0              &  0 \\
0              & 0             & -|J_2|     & 0        & \Sg       & \Delta_A & 0              & -J_1 \\
0              & 0             & 0          & |J_2|    & -\Delta_A & -\Sg     & J_1            & 0 \\ 
0              & 0             & 0          & 0        & 0         & -J_1     & \Sg            & \Delta_A \\ 
0              & 0             & 0          & 0        & J_1       & 0        & -\Delta_A      & -\Sg
\end{array}
\right]
\label{eq:D_H_Gs}
\ee
and
\begin{equation}
\tv =
\left[
\begin{array}{cccccccc} 
0 & 0 & 0 & 0 & 0 & 0 & 0 & 0 \\
0 & 0 & 0 & 0 & 0 & 0 & 0 & 0 \\
0 & 0 & 0 & 0 & 0 & 0 & 0 & 0 \\
0 & 0 & 0 & 0 & 0 & 0 & 0 & 0 \\
0 & 0 & 0 & 0 & 0 & 0 & 0 & 0 \\
0 & 0 & 0 & 0 & 0 & 0 & 0 & 0 \\
-|J_2| & 0 & 0 & 0 & 0 & 0 & 0 & 0 \\
0 & |J_2| & 0 & 0 & 0 & 0 & 0 & 0 
\end{array}
\right] ,
\label{eq:t_H_Gs}
\end{equation}
and $\Iv$ is the $8\times8$ identity matrix.
With these definitions of $\Dv$ and $\tv$, 
we calculate the surface Green's function ($G^s$) of the semi-infinite chain terminated on the left with atom $A$
using the decimation algorithm \cite{Sancho_Rubio_JPhysF84,MPLSancho_HighlyConvergent_JPF85,Galperin:JCP:2002:decimation}
described in Appendix \ref{app:decimation}.
The surface spectral functions, given by $A_s = -2 {\rm Im}\left(G^s_{1,1} \right)$,
calculated for the same parameters used in the $E-k$ plots of 
Figs. \ref{fig:E-k_doubly_dimerized_afm} and \ref{fig:2_dimer_E-ks_3}
are shown in Fig. \ref{fig:2_dimer_As-As_3}.
The surface state appears for $J_2 > J_{2c}$ and disappears for  $J_2 < J_{2c}$.
Thus, based on the bulk-boundary correspondence, the magnon spectrum is in a 
topologically non-trivial state for $J_2 > J_{2c}$.

Since this is a 1D chain, it is tempting to view it as a variant of the SSH model, however, this system is 
qualitatively different.
In the original electronic version of the SSH model \cite{1979_SSH_PRL}, bandgap closing occurs 
at X (the zone boundary) when the
tight-binding hopping matrix elements within the unit cell $(t_1)$ and between unit cells $(t_2)$ are equal,
i.e. $t_1 = t_2$. 
Furthermore, the topology is trivial for $|t_1| > |t_2|$ and non-trivial, with a Zak phase equal to 1, when $|t_1| < |t_2|$. 
Finally, the existence of a surface state in the middle of the bulk gap follows the Shockley criterion,
in which a surface state exists when the bond of larger magnitude is cut \cite{1939_Surface_States_Shockley}.
Dimerized FM chains, also follow this model.
In a dimerized FM chain with alternating values for the nearest neighbor exchange couplings,
the magnon gap closes at X when the two exchange constants are equal \cite{2022_SSH_Magnons_JPCM}.
In a dimerized chain of FM spheres with nearest-neighbor dipolar coupling, the magnon gap also closes at X
when the dipolar coupling within and between unit cells is equal \cite{2018_SSH_Magnon_Chain}.
For both of the above systems, a surface state exists when the linkage corresponding to the stronger coupling is cut.
Our 1D chain is doubly dimerized, i.e. there are 4 atoms per magnetic unit cell.
Thus, trivial zone-folding occurs such that the Y-point is folded back to $\Gamma$ and the gap now closes at $\Gamma$.
What is qualitatively different about our system is that the magnon gap does not close when $|J_2| = J_1$, 
but when $|J_2| \approx A_y/2$ (the easy plane anisotropy).
For large $J_1$, Eq. (\ref{eq:Jcrit}) shows that the critical value for $J_2$ has negligible dependence on $J_1$. 
Finally, the surface state in the gap exists when we cut the {\em weaker} coupling $J_2$ rather than the stronger coupling $J_1$.
Thus, the physics governing the topological properties this 1D magnonic system are qualitatively different 
from the physics of the SSH model.
\begin{figure}[t]
\centering
\includegraphics[width=3.2in]{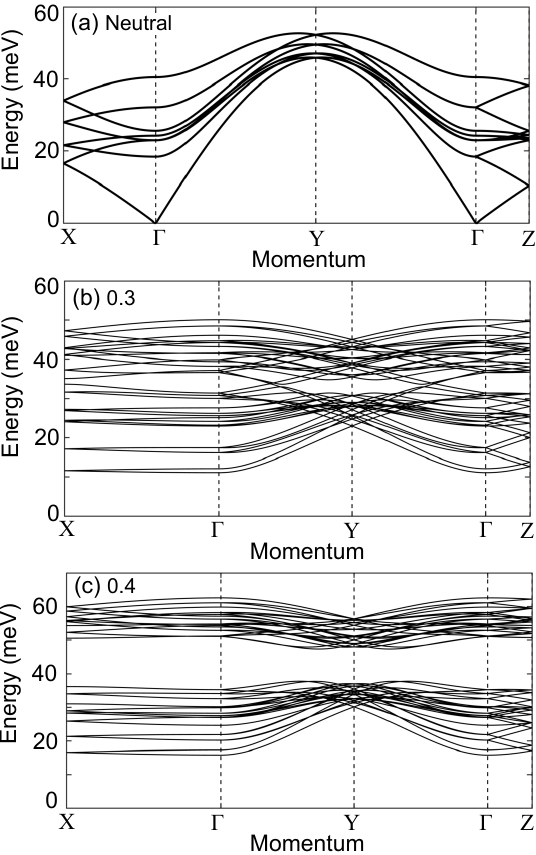}
\caption{
Bulk magnon dispersions calculated for (a) neutral, (b) 0.3, and (c) 0.4 hole filling.
All values are taken from the Table \ref{tab:parameters_filling}.
}
\label{fig:Filling_0_3_4_Bulk_E-ks}
\end{figure}
\section{Interchain Coupling}
\label{sec:Interchain_Coupling}

\subsection{AFM Charge Neutral Bulk Structure}
\label{sec:Charge_Neutral_Structure}
In the bulk, the chains are arranged in a triangular lattice in the $a-c$ plane
as shown in Fig. \ref{fig:MoI3_struct_0}(c,d). 
The AFM interchain couplings ($J_3$ and $J_4$) combined with the triangular lattice give rise to spin frustration,
and the lowest energy ground-state spin texture that we have found
for the charge neutral bulk structure,
both from DFT total energy calculations and from the spin Hamiltonian, is a helical texture,
shown in Fig. \ref{fig:MoI3_struct_0}(c,d), such that the
interchain, nearest-neighbor spins are rotated by $120^\circ$ with respect to each other.
A further indication that this state is a stable energy minimum is that the magnon dispersion
calculations using this state as the gound state give no negative modes. 
This is similar to the criterion used to determine a stable crystal structure by calculating the phonon
dispersion and finding no negative modes.
If, for example, we attempt to calculate the magnon dispersion of the charge neutral structure
starting from a collinear state, we observe many negative modes.

The bulk magnon dispersion of the charge neutral structure 
is shown in Fig. \ref{fig:Filling_0_3_4_Bulk_E-ks}(a).
Even though the interchain exchange couplings ($J_3$ and $J_4$) 
are relatively small compared to $J_1$ and $J_2$, each chain is coupled to
6 neighbors, which significantly enhances their effect.
The AFM interchain coupling causes splitting of the single-chain dispersion along the chain direction $(\Gamma-Y)$,
and it gives rise to cross-chain dispersion ($\Gamma-X$ and $\Gamma-Z$). 
The energies of the first two excited states at $\Gamma$ of 18.4 meV and 23.0 meV 
straddle the energy of the first excited state of the isolated 
chain of 22.0 meV.
\subsection{AFM / FM Bulk Structure}
\label{sec:AFM_FM_Bulk}
Considering the parameters in Table \ref{tab:parameters_filling} and Eq. (\ref{eq:Jcrit}), 
the topological transition occurs between
a hole doping of 0.2 and 0.3 per primitive unit cell (16 atoms), 
and the structures with 0.3 and 0.4 hole doping are in the topological magnonic state.
The magnon dispersions calculated for 0.3 and 0.4 hole filling are shown in Figs. 
\ref{fig:Filling_0_3_4_Bulk_E-ks}(b,c).
For the smaller value of $|J_2| = 2.30$ meV, the gap in the dispersion is closed by the
splitting resulting from the interchain coupling.
For the larger value of $|J_2| = 4.26$ meV, the gap in the spectrum of the
single-chain spectrum is large enough that the gap remains open in the presence of the
interchain coupling, and the topological character of the single-chain dispersion
can still be observed.
A second result of increased hole doping is increased in-plane ($a-c$ plane) anisotropy ($A_x$). 
For both 0.3 and 0.4 hole doping,
the ground state spin spiral texture of the charge neutral system shown in Fig. \ref{fig:MoI3_struct_0}(c,d) 
is replaced by the collinear spin texture with all spins aligned 
along the $\pm c$ axis ($\pm z$).
This minimizes the anisotropy energy from the last term in the Hamiltonian in Eq. (\ref{eq:doubly_dimerized_AFM_H}).
\section{Conclusions}
\label{sec:conclusions}
Alternating AFM ($J_1$) and FM ($J_2$) exchange couplings along a spin chain can give rise to 1D topological
magnon bands depending on the relative magnitudes of $J_1$, $J_2$, and the anisotropy constants.
The topological phase is qualitatively different from that of the SSH model.
The existence of such a phase requires dimerization ($|J_2| \neq |J_1|$), 
alternating AFM and FM coupling ($J_1>0$, $J_2 < 0$), and easy-plane anisotropy.
The condition for the topological phase with large $J_1$ is $|J_2| \gtrsim A_y/2$.
This is a qualitatively different condition than found in the SSH model.
Furthermore, the surface state exists when the weaker bond corresponding to the weaker coupling ($J_2$) is cut.
This condition also contradicts the SSH model.
This model system may be physically embodied in MoI$_3$.
In bulk MoI$_3$, the small AFM coupling between the chains, when multiplied by the 6 nearest neighbor chains,
becomes significant.
Both the sign and magnitude of the intrachain exchange coupling along the longer bond ($J_2$)
are affected by hole filling. 
The in-plane anisotropy is also affected by hole filling.
At larger values of hole filling, the single-chain, topological magnonic gap remains larger than the
band splitting resulting from the cross-chain coupling.
As a result, MoI$_3$ may provide a material platform in which
multiple magnetic phases, both normal and topological, can be obtained by gating or doping.

\clearpage
\begin{acknowledgments}
This work was supported by the Vannevar Bush Faculty Fellowship from the Office of Secretary of Defense (OSD) 
under the Office of Naval Research (ONR) contract N00014-21-1-2947 on One-Dimensional Quantum Materials.
DFT calculations were performed on STAMPEDE2 at TACC and EXPANSE at SDSC under allocation DMR130081 from the 
Advanced Cyberinfrastructure Coordination Ecosystem: Services $\&$ Support (ACCESS) program \cite{ACCESS}, 
which is supported by National Science Foundation grants $\#$2138259, $\#$2138286, $\#$2138307, $\#$2137603, and $\#$2138296.
R. Lake acknowledges useful discussions with Ran Cheng and Hantao Zhang.
\end{acknowledgments}

\appendix
\section{Exchange constants}
\label{app:exchange}

\begin{figure*}[tb]
\centering
\includegraphics[width=6.5in]{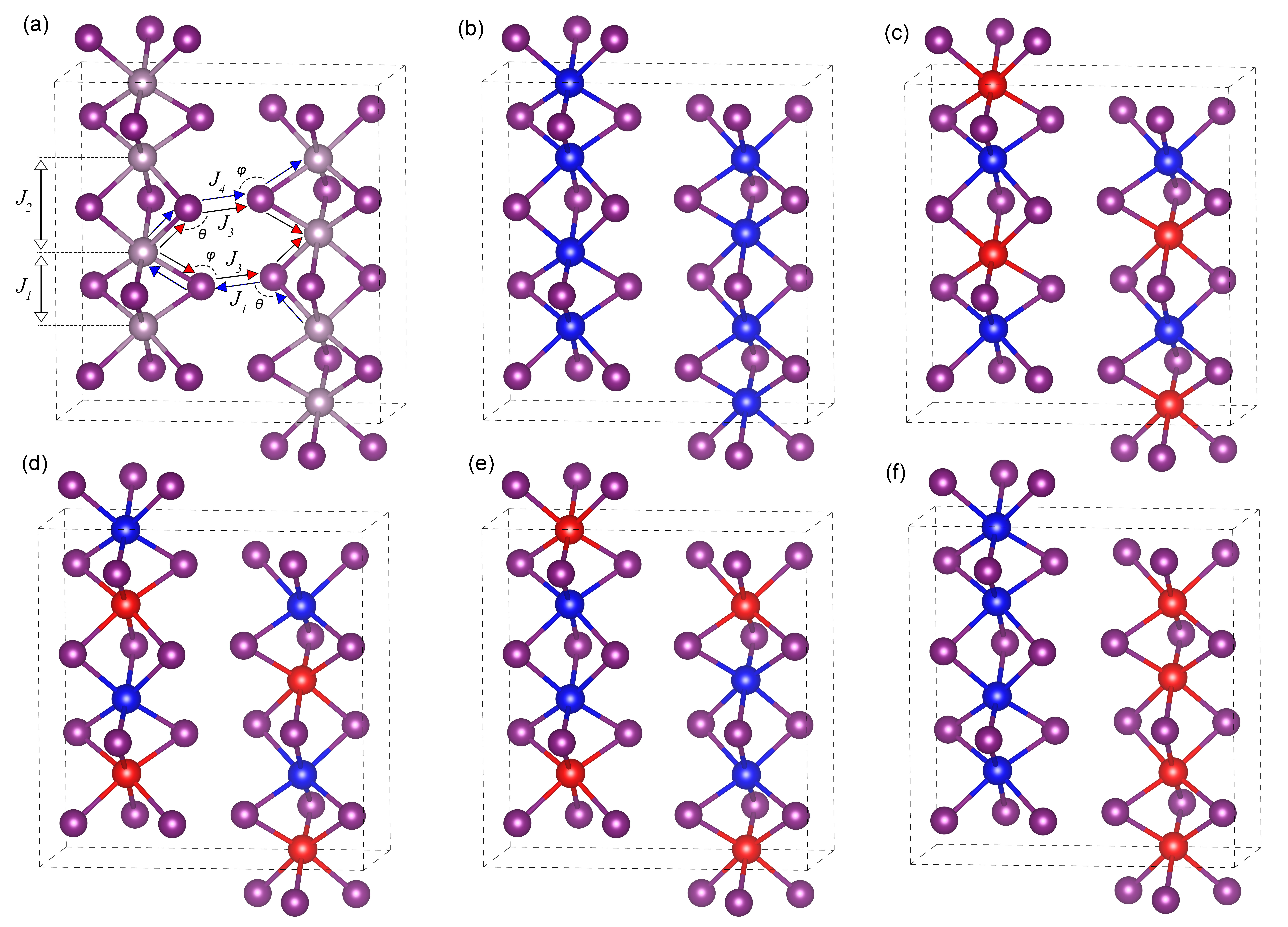}
\caption{
(a) Illustration of the 4 exchange constants. 
The shortest through-bond paths are shown for $J_3$ (in red) and $J_4$
(in blue).
The bond angles for the upper and lower paths for $J_4$ are shown. 
The angles are $\theta = 91.6^\circ$ and $\varphi = 139.2^\circ$. 
The lower path is the reverse of the upper path.
(b-f) Different spin configurations used for the total energy calculations.
Red and blue atoms represent opposite spins.
}
\label{fig:app_J}
\end{figure*}

To calculate the exchange constants, we use the energy mapping method
\cite{2009_Whangbo_diamon_chain_JPCM,2022_Na2Cu2TeO6_Dagatto_PRB,2022_CrTe2_YLiu_PRMat,2022_MoI3_Fari_APL}.
In this method, we calculate the total energy for different FM and AFM configurations of \MoI 
using PBE-D3 with spin-orbit coupling (SOC) and equate this energy with the 
energy resulting from the Hamiltonians in Eqs. (\ref{eq:dimerized_AFM_H}) or (\ref{eq:doubly_dimerized_AFM_H}).
Since there are 4 exchange constants ($J_1$ ... $J_4$) and an unknown energy $E_0$
resulting from the non-magnetic components of the total energy functional in the DFT calculations,
we consider five different spin configurations (1 FM and 4 AFM states), as illustrated
in Fig. \ref{fig:app_J}.
All spins are aligned along the $\pm c$ axis.
In all of these configurations, the unit cell is doubled along the chain direction.
As shown in figure \ref{fig:app_J}(a), the exchange constants J$_1$ and $J_2$ represent 
the intrachain short and long Mo-Mo coupling, 
and $J_3$ and $J_4$ are the interchain exchange couplings.
The primary through-bond paths mediating $J_3$ and $J_4$ are illustrated in Fig. \ref{fig:app_J}(a).
The total energy of these five configurations can be derived from the magnetic Hamiltonian as:
\begin{equation}
\begin{aligned}
E_{FM} &= E_0 + 8J_1 + 8J_2 + 48J_3 + 96J_4 \\
E_{AFM1} &= E_0 - 8J_1 - 8J_2 + 48J_3 - 96J_4 \\
E_{AFM2} &= E_0 - 8J_1 - 8J_2 - 16J_3 + 32J_4 \\
E_{AFM3} &= E_0 - 8J_1 + 8J_2 + 16J_3 + 0J_4 \\
E_{AFM4} &= E_0 + 8J_1 + 8J_2 - 16J_3 - 32J_4
\end{aligned}
\label{eq:exch_J}
\end{equation}

To determine the anisotropy energies, the spin configuration of
Fig. \ref{fig:app_J}(d) is used.
Three total energy calculations are performed with all spins aligned along
$\pm x$, $\pm y$, and $\pm z$, and the energy differences
give the anisotropy energies per Mo atom,
$A_x = ( E_x - E_z) / 8$ 
and 
$A_y = ( E_y - E_z ) / 8$,
where $E_\nu$ is the total energy of the supercell with the spins aligned along
the $\pm \nu$ directions.

The above energy mapping approach, while heavily used throughout the literature, can be viewed as the lowest level
of theory for constructing a spin Hamiltonian that maps the total energies
calculated from DFT. 
At the next level of sophistication, 
the exchange constants can be replaced by exchange tensors
as described in 
\cite{2020_Generic_4_State_E_Map_Jij_PRB,2021_2D_FMs_Vandenberghe_PRB}.
The tensors representing $J_1$ and $J_2$ in \MoI each have 
three independent symmetry allowed diagonal elements.
The symmetry allowed full symmetric tensors representing $J_3$ and $J_4$ 
each have 6 independent elements.
Thus, the symmetry allowed exchange tensors representing 
$J_1$-$J_4$ contain 18 exchange parameters.
Using the four-state energy mapping method \cite{2020_Generic_4_State_E_Map_Jij_PRB}
would require a $3\times 3 \times 3$ supercell, consisting of 432 atoms,
and 4 total energy calculations for each of the 18 exchange parameters.
The number of parameters can be reduced by using analytical expressions for the
exchange constants and assuming symmetries to reduce the number of 
free parameters, such as in-plane isotropy \cite{2021_2D_FMs_Vandenberghe_PRB}. 
Determining the full tensors would require a more sophisticated approach such as
the Green's function or Liechtenstein approach 
\cite{2015_Green_Fn_Calc_J_PRB,2019_Liechtenstein_Method_JPSJ,2021_TB2J_J_Greens_fn_CPC}.
These latter approaches can provide greater physical insight into the exchange couplings resulting from
different orbitals, however, when using a plane-wave DFT code, they require a transformation into the Wannier basis,
which can be problematic for larger unit cells. 

Under hole doping, the presence of itinerant holes raises questions about the validity of the
Heisenberg type Hamiltonians, which assume localized spins, relatively localized exchange interactions,
and absence of longitudinal spin modes (i.e. uniform magnitude of magnetic moments).
Electronic structure calculations described in App. \ref{app:ek} show that the 
partially occupied frontier d-orbital valence band is 
very narrow indicating strong localization of the holes residing on the Mo atoms.
Furthmore the self-consistent field DFT calculations of the supercells find that the magnitudes of 
the Mo magnetic moments to be identical.
Thus, the use of the Heisenberg type Hamiltonians can be justified.

\section{Electronic bands}
\label{app:ek}

\begin{figure*}[tb]
\centering
\includegraphics[width=7in]{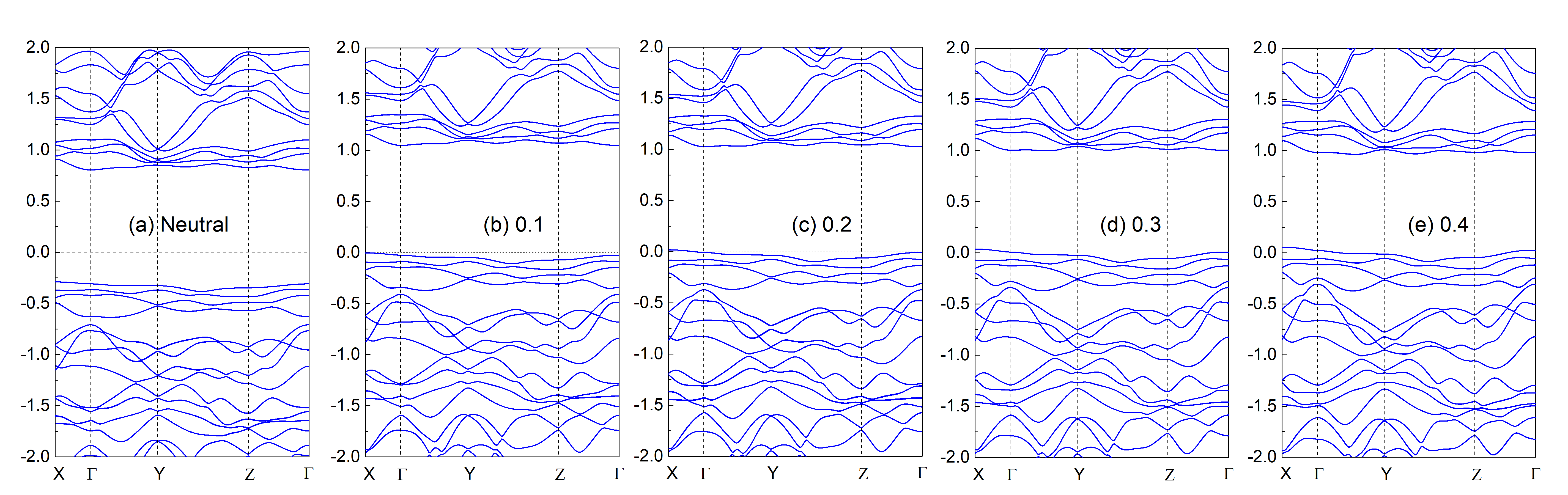}
\caption{Electronic band diagrams of \MoI for different hole fillings 
as labeled in the plots.}
\label{fig:appendix-ek}
\end{figure*}

To see the effect of hole filling on the Fermi level, 
we have calculated the electronic bands of \MoI with SOC.
Figure \ref{fig:appendix-ek} shows the electronic bands of the AFM state of \MoI with different hole fillings. 
Neutral \MoI is a semiconductor with a calculated bandgap of 1.094 eV 
(which is close to the experimental value of 1.27 eV \cite{MoI3_exp}).
The uppermost 4 valence bands are narrow bands composed of the d-orbitals of the Mo atoms.
Hole doping causes the uppermost valence band to become partially unoccupied.
The narrow bandwidth of the uppermost valence band indicates strong localization of the frontier
d-orbital band and thus strong localization of the itinerant holes.

\section{Decimation Algorithm}
\label{app:decimation}
Using matrices (\ref{eq:D_H_Gs}) and (\ref{eq:t_H_Gs}), we implement the decimation algorithm 
\cite{Sancho_Rubio_JPhysF84,MPLSancho_HighlyConvergent_JPF85,Galperin:JCP:2002:decimation}
for calculating the surface Green function.
For calculating the surface Green function of the semi-infinite slab terminated on the left with atom $A$,
we define the following
\begin{equation}
\begin{aligned}
\Dv^s_0&= \Dv\\
\Dv_0&= \Dv\\
\Av_0&= \tv\\
\Bv_0&= \tv^\dagger ,
\end{aligned}
\label{eq:Decimation_0}
\end{equation}
where the subscript now indicates the iteration number.
We now iterate 
\begin{equation}
\begin{aligned}
\Dv_i^s =& \: \Dv_{i-1}^s + \Av_{i-1} \left[ (E + i\eta)\Iv - \Dv_{i-1} \right]^{-1} \Bv_{i-1} \\
\Dv_i =& \: \Dv_{i-1} + \Av_{i-1} \left[ (E + i\eta)\Iv - \Dv_{i-1} \right]^{-1} \Bv_{i-1} \\
& + \Bv_{i-1} \left[ (E + i\eta)\Iv - \Dv_{i-1} \right]^{-1} \Av_{i-1} \\
\Av_i =& \: \Av_{i-1} \left[ (E + i\eta)\Iv - \Dv_{i-1} \right] \Av_{i-1} \\
\Bv_i =& \: \Bv_{i-1} \left[ (E + i\eta)\Iv - \Dv_{i-1} \right] \Bv_{i-1}
\end{aligned}
\label{eq:Decimation_iter}
\end{equation}
until the original $2\times2$ non-zero coupling blocks of $\Av_i$ are negligible. 
In practice, once all elements of $\Av_i$ are less than $1.0 \times 10^{-9}$ meV, we exit the iteration loop.
The $8\times8$ surface Green function is then
\begin{equation}
\Gv^s = \left( E \Iv - \Dv_i^s \right)^{-1} .
\label{eq:Decimation_converged}
\end{equation}
The value of $\eta$ used was $1.0 \times 10^{-3}$ meV.

\section{Effective Field Calculation of Magnon Dispersion}
\label{app:Eff_Field_Magnon}
In this appendix, we describe the classical, effective field approach for calculating the magnon
dispersions for the single-chain structures illustrated in Fig. \ref{1dimerized_2dimerized_chain}.

\subsection{Dimerized Chain with $J_1, J_2 > 0$}
\label{app:singly_dimerized_eff_field}
We first consider the structure of Fig. \ref{1dimerized_2dimerized_chain}(a) with the Hamiltonian 
given in Eq. (\ref{eq:dimerized_AFM_H}).
From the Hamiltonian, we identify the effective field acting on the the spins.
The magnetic moment associated with each spin $A$ or $B$ in the unit cell is
\be
m_n^{A,B} = -g \mu_B \Sv_n^{A,B}
\ee
The effective field acting on magnetic moment $m_n^\alpha$ is given by $-\partial H/\partial m_n^\alpha$.
The effective fields are
\be
\Bv_{{\rm eff},n}^A =
\frac{2}{g\mu_B} \left( J_1 \Sv_n^B + J_2 \Sv_{n-1}^B \right) +
\frac{2A_y}{g\mu_B} S_{n,y}^A \yh +
\frac{2A_x}{g\mu_B} S_{n,x}^A \xh ,
\nonumber
\ee
and
\be
\Bv_{{\rm eff},n}^B =
\frac{2}{g\mu_B} \left( J_1 \Sv_n^A + J_2 \Sv_{n+1}^A \right) +
\frac{2A_y}{g\mu_B} S_{n,y}^B \yh +
\frac{2A_x}{g\mu_B} S_{n,x}^B \xh .
\nonumber
\ee
The equations of motions for the angular momenta defined by the spins are
\be
\hbar \frac{d\Sv_n^A}{dt} = \muv_n^A \times \Bv_{{\rm eff},n}^A
\ee
or 
\be
\hbar \frac{d\Sv_n^A}{dt} = -g\mu_B \Sv_n^A \times \Bv_{{\rm eff},n}^A,
\label{eq:eom_SA_BA}
\ee
and similarly,
\be
\hbar \frac{d\Sv_n^B}{dt} = -g\mu_B \Sv_n^B \times \Bv_{{\rm eff},n}^B .
\label{eq:eom_SB_BB}
\ee
Writing out (\ref{eq:eom_SA_BA}) and (\ref{eq:eom_SB_BB}) gives
\begin{align}
\hbar \dot{\Sv}_n^A & = - \Sv_n^A \times 2\left[
J_1 \Sv_n^B + J_2 \Sv_{n-1}^B  + A_y S_{n,y}^A \yh + A_x S_{n,x}^A \xh 
\right]
\nonumber \\
\hbar \dot{\Sv}_n^B & = - \Sv_n^B \times 2\left[
J_1 \Sv_n^A + J_2 \Sv_{n+1}^A + A_y S_{n,y}^B \yh + A_x S_{n,x}^B \xh
\right] .
\nonumber
\end{align}
In linear response, we assume the deviations from the equilibrium alignments are small.
Therefore, $S_{n,z}^A = S$ and $S_{n,z}^B = -S$.
The equations of motion for the $x$ and $y$ components of $\dot{\Sv}_n^A$ and $\dot{\Sv}_n^B$ are
\begin{equation}
\begin{aligned}
\hbar \dot{S}_{n,x}^A =&
2S \left[ (J_1 + J_2 + A_y ) S^A_{n,y} + J_1  S^B_{n,y} + J_2 S^B_{n-1,y} \right] 
\\
\hbar \dot{S}_{n,y}^A  =&
- 2S \left[ (J_1 + J_2 + A_x ) S^A_{n,x} + J_1  S^B_{n,x} + J_2 S^B_{n-1,x} \right] 
\\
\hbar \dot{S}_{n,x}^B =&
- 2S \left[ (J_1 + J_2 + A_y ) S^B_{n,y} + J_1  S^A_{n,y} + J_2 S^A_{n+1,y} \right]  
\\
\hbar \dot{S}_{n,y}^B =&
2S \left[ (J_1 + J_2 + A_x ) S^B_{n,x} + J_1  S^A_{n,x} + J_2 S^A_{n+1,x} \right] .
\label{eq:dSABxy_dt_ssh_afm_classical}
\end{aligned}
\end{equation}
Assume a plane-wave solution such that
\begin{align}
S^A_{n,x} = S^A_x e^{i(kbn - \w t)}
\nonumber \\
S^A_{n,y} = S^A_y e^{i(kbn - \w t)}
\nonumber \\
S^B_{n,x} = S^B_x e^{i(kbn - \w t)}
\nonumber \\
S^B_{n,y} = S^B_y e^{i(kbn - \w t)}
\end{align}
where the unknown coefficients $S^A_x$, $S^A_y$, $S^B_x$, and $S^B_y$ are complex amplitudes,
$b$ is the lattice constant, and $n$ is the unit cell index.
Placing these forms of the solutions into Eqs. (\ref{eq:dSABxy_dt_ssh_afm_classical}), we have
\begin{align}
-i \hbar \w S^A_x &= 2S \left[ (J_1 + J_2 + A_y ) S^A_y + \left( J_1  + J_2 e^{-ika} \right)   S^B_y  \right]
\nonumber \\
-i \hbar \w S^A_y &= -2S \left[ (J_1 + J_2 + A_x ) S^A_x + \left( J_1  + J_2 e^{-ika} \right)   S^B_x  \right]
\nonumber \\
-i \hbar \w S^B_x &= -2S \left[ (J_1 + J_2 + A_y ) S^B_y + \left( J_1  + J_2 e^{+ika} \right)   S^A_y  \right]
\nonumber \\
-i \hbar \w S^B_y &= 2S \left[ (J_1 + J_2 + A_x ) S^B_x + \left( J_1  + J_2 e^{+ika} \right)   S^A_x  \right]  .
\label{eq:EOM_classical_w_k}
\end{align}
Re-writing (\ref{eq:EOM_classical_w_k}) as a matrix eigenvalue equation gives
\begin{widetext}
\begin{equation}
\left[
\begin{array}{cccc}
-\hbar \w & i 2S (J_1 + J_2 + A_y ) & 0 & i2S \left( J_1  + J_2 e^{-ika} \right) \\
-i 2S (J_1 + J_2 + A_x ) & -\hbar \w & -i 2S \left( J_1  + J_2 e^{-ika} \right) & 0 \\
0 & -i 2S \left( J_1  + J_2 e^{+ika} \right) & -\hbar \w & -i 2S (J_1 + J_2 + A_y ) \\
i 2S \left( J_1  + J_2 e^{+ika} \right) & 0 & i 2S (J_1 + J_2 + A_x ) & -\hbar \w \\
\end{array}
\right]
\left[
\begin{array}{c}
S^A_x\\
S^A_y\\
S^B_x\\
S^B_y
\end{array}
\right]
=
0 .
\end{equation}
\end{widetext}
Setting the determinant to zero, gives 4 solutions for $\hbar\w$ 
of the form $\pm \hbar\w_0(k)$ and $\pm \hbar\w_1(k)$.  
The two positive solutions are given by Eqs. (\ref{eq:w0-k_1D_dimerized_AFM}) and 
(\ref{eq:w1-k_1D_dimerized_AFM}) of the main text.

\subsection{Tetramerized Chain with $J_1 > 0$ and  $J_2 < 0$}
\label{app:doubly_dimerized_eff_field}
We now consider the structure in Fig. \ref{1dimerized_2dimerized_chain}(b) 
with Hamiltonian given by Eq. (\ref{eq:doubly_dimerized_AFM_H}).
Proceeding as above, the effective fields are
\begin{align}
B_{{\rm eff},n}^A & = \tfrac{2}{g\mu_B} 
\left[ J_1 S_n^B + J_2 S^D_{n-1} + A_y S^A_{n,y} \hat{y} + A_x S^A_{n,x} \hat{x} \right]
\nonumber \\
B_{{\rm eff},n}^B & = \tfrac{2}{g\mu_B}
\left[ J_1 S_n^A + J_2 S^C_{n} + A_y S^B_{n,y} \hat{y} + A_x S^B_{n,x} \hat{x} \right]
\nonumber \\
B_{{\rm eff},n}^C & = \tfrac{2}{g\mu_B}
\left[ J_2 S_n^B + J_1 S^D_{n} + A_y S^C_{n,y} \hat{y} + A_x S^C_{n,x} \hat{x} \right]
\nonumber \\
B_{{\rm eff},n}^D & = \tfrac{2}{g\mu_B}
\left[ J_1 S_n^C + J_2 S^A_{n+1} + A_y S^D_{n,y} \hat{y} + A_x S^D_{n,x} \hat{x} \right] ,
\label{eq:Beff_doubly_dimerized_afm}
\end{align}
and the resulting equations of motion are
\begin{align}
\hbar \dot{S}_n^A  & = -S_n^A \times 2 
\left[ J_1 S_n^B + J_2 S^D_{n-1} + A_y S^A_{n,y} \hat{y} + A_x S^A_{n,x} \hat{x} \right]
\nonumber \\
\hbar \dot{S}_n^B  & = -S_n^B \times 2 
\left[ J_1 S_n^A + J_2 S^C_{n} + A_y S^B_{n,y} \hat{y} + A_x S^B_{n,x} \hat{x} \right]
\nonumber \\
\hbar \dot{S}_n^C  & = -S_n^C \times 2 
\left[ J_2 S_n^B + J_1 S^D_{n} + A_y S^C_{n,y} \hat{y} + A_x S^C_{n,x} \hat{x} \right]
\nonumber \\
\hbar \dot{S}_n^D  & = -S_n^D \times 2 
\left[ J_1 S_n^C + J_2 S^A_{n+1} + A_y S^D_{n,y} \hat{y} + A_x S^D_{n,x} \hat{x} \right] .
\label{eq:eoms_doubly_dimerized_afm}
\end{align}
Inserting plane wave solutions $S^\alpha_{n,xy} = S^\alpha_{xy} e^{i(kan - \omega t)}$
and setting $S^A_z = S^D_z = S$ and $S^B_z = S^C_z = -S$, we have
\[
\hbar \omega S^A_x =   2iS \left[ (J_1 - J_2 + A_y) S_y^A + J_1 S_y^B + J_2 S_y^D e^{-ika} \right]
\]
\[
\hbar \omega S^A_y = - 2iS \left[ (J_1 - J_2 + A_x) S_x^A + J_1 S_x^B + J_2 S_x^D e^{-ika} \right]
\]
\[
\hbar \omega S^B_x = - 2iS \left[ (J_1 - J_2 + A_y) S_y^B + J_1 S_y^A + J_2 S_y^C \right]
\]
\[
\hbar \omega S^B_y =   2iS \left[ (J_1 - J_2 + A_x) S_x^B + J_1 S_x^A + J_2 S_x^C  \right]
\]
\[
\hbar \omega S^C_x = - 2iS \left[ (J_1 - J_2 + A_y) S_y^C + J_2 S_y^B + J_1 S_y^D \right]
\]
\[
\hbar \omega S^C_y =   2iS \left[ (J_1 - J_2 + A_x) S_x^C + J_2 S_x^B + J_1 S_x^D  \right]
\]
\[
\hbar \omega S^D_x =   2iS \left[ (J_1 - J_2 + A_y) S_y^D + J_1 S_y^C + J_2 S_y^A e^{ika} \right]
\]
\[
\hbar \omega S^D_y = - 2iS \left[ (J_1 - J_2 + A_x) S_x^D + J_1 S_x^C + J_2 S_x^A  e^{ika} \right]
\]
Using the following definitions,
\begin{align}
t_{x} = 2iS (J_1 - J_2 + A_x)
\nonumber \\
t_{y} = 2iS (J_1 - J_2 + A_y)
\nonumber \\
\tJl = 2iS J_1
\nonumber \\
\tJtwo = 2iS J_2,
\label{eq:t_J_defs}
\end{align}
the energies $\hbar \omega$ are the eigenenergies of an $8\times8$ matrix,
\begin{widetext}
\begin{equation}
\left[
\begin{array}{cccccccc} 
-\hw & t_{y} & 0 & \tJl & 0 & 0 & 0 & \tJtwo e^{-ikb_m} \\
-t_{x} & -\hw & \tJl & 0 & 0 & 0 & -\tJtwo e^{-ikb_m} & 0\\
0 & -\tJl & -\hw & -t_y & 0 & -\tJtwo & 0 & 0\\
\tJl & 0 & t_x & -\hw & \tJtwo & 0 & 0 & 0\\
0 & 0 & 0 & -\tJtwo & -\hw & -t_y & 0 & -\tJl\\
0 & 0 & \tJtwo & 0 & t_x & -\hw & \tJl & 0\\
0 & \tJtwo e^{ikb_m} & 0 & 0 & 0 & \tJl & -\hw & t_y\\
-\tJtwo e^{ikb_m} & 0 & 0 & 0 & -\tJl & 0 & -t_x & -\hw
\end{array}
\right]
\left[
\begin{array}{c}
S_x^A\\
S_y^A\\
S_x^B\\
S_y^B\\
S_x^C\\
S_y^C\\
S_x^D\\
S_y^D
\end{array}
\right]
= 0 .
\label{eq:eig_eqn_doubly_deg_afm}
\end{equation}
Setting the determinant to zero gives 4 pairs of bands $\pm \hw(k)$.
The four positive bands are given by Eqs. (\ref{eq:E-k_1,2_dgafm}) and (\ref{eq:E-k_3,4_dgafm}) of the main text.
\end{widetext}

\providecommand{\noopsort}[1]{}\providecommand{\singleletter}[1]{#1}%


\begin{thebibliography}{43}%
\makeatletter
\providecommand \@ifxundefined [1]{%
 \@ifx{#1\undefined}
}%
\providecommand \@ifnum [1]{%
 \ifnum #1\expandafter \@firstoftwo
 \else \expandafter \@secondoftwo
 \fi
}%
\providecommand \@ifx [1]{%
 \ifx #1\expandafter \@firstoftwo
 \else \expandafter \@secondoftwo
 \fi
}%
\providecommand \natexlab [1]{#1}%
\providecommand \enquote  [1]{``#1''}%
\providecommand \bibnamefont  [1]{#1}%
\providecommand \bibfnamefont [1]{#1}%
\providecommand \citenamefont [1]{#1}%
\providecommand \href@noop [0]{\@secondoftwo}%
\providecommand \href [0]{\begingroup \@sanitize@url \@href}%
\providecommand \@href[1]{\@@startlink{#1}\@@href}%
\providecommand \@@href[1]{\endgroup#1\@@endlink}%
\providecommand \@sanitize@url [0]{\catcode `\\12\catcode `\$12\catcode
  `\&12\catcode `\#12\catcode `\^12\catcode `\_12\catcode `\%12\relax}%
\providecommand \@@startlink[1]{}%
\providecommand \@@endlink[0]{}%
\providecommand \url  [0]{\begingroup\@sanitize@url \@url }%
\providecommand \@url [1]{\endgroup\@href {#1}{\urlprefix }}%
\providecommand \urlprefix  [0]{URL }%
\providecommand \Eprint [0]{\href }%
\providecommand \doibase [0]{https://doi.org/}%
\providecommand \selectlanguage [0]{\@gobble}%
\providecommand \bibinfo  [0]{\@secondoftwo}%
\providecommand \bibfield  [0]{\@secondoftwo}%
\providecommand \translation [1]{[#1]}%
\providecommand \BibitemOpen [0]{}%
\providecommand \bibitemStop [0]{}%
\providecommand \bibitemNoStop [0]{.\EOS\space}%
\providecommand \EOS [0]{\spacefactor3000\relax}%
\providecommand \BibitemShut  [1]{\csname bibitem#1\endcsname}%
\let\auto@bib@innerbib\@empty
%
\bibitem [{\citenamefont {Wang}\ \emph {et~al.}(2018)\citenamefont {Wang},
  \citenamefont {Zhang},\ and\ \citenamefont
  {Wang}}]{2018_Topo_Magnonics_Paradigm_PRApp}%
  \BibitemOpen
  \bibfield  {author} {\bibinfo {author} {\bibfnamefont {X.~S.}\ \bibnamefont
  {Wang}}, \bibinfo {author} {\bibfnamefont {H.~W.}\ \bibnamefont {Zhang}},\
  and\ \bibinfo {author} {\bibfnamefont {X.~R.}\ \bibnamefont {Wang}},\
  }\bibfield  {title} {\bibinfo {title} {Topological magnonics: A paradigm for
  spin-wave manipulation and device design},\ }\href
  {https://doi.org/10.1103/PhysRevApplied.9.024029} {\bibfield  {journal}
  {\bibinfo  {journal} {Phys. Rev. Appl.}\ }\textbf {\bibinfo {volume} {9}},\
  \bibinfo {pages} {024029} (\bibinfo {year} {2018})}\BibitemShut {NoStop}%
\bibitem [{\citenamefont {Li}\ \emph {et~al.}(2021)\citenamefont {Li},
  \citenamefont {Cao},\ and\ \citenamefont
  {Yan}}]{2021_Topo_Magnon_Review_PhysRep}%
  \BibitemOpen
  \bibfield  {author} {\bibinfo {author} {\bibfnamefont {Z.-X.}\ \bibnamefont
  {Li}}, \bibinfo {author} {\bibfnamefont {Y.}~\bibnamefont {Cao}},\ and\
  \bibinfo {author} {\bibfnamefont {P.}~\bibnamefont {Yan}},\ }\bibfield
  {title} {\bibinfo {title} {Topological insulators and semimetals in classical
  magnetic systems},\ }\href {https://doi.org/10.1016/j.physrep.2021.02.003}
  {\bibfield  {journal} {\bibinfo  {journal} {Physics Reports}\ }\textbf
  {\bibinfo {volume} {915}},\ \bibinfo {pages} {1} (\bibinfo {year}
  {2021})}\BibitemShut {NoStop}%
\bibitem [{\citenamefont {McClarty}(2022)}]{2022_Topo_Magnons_review_McClarty}%
  \BibitemOpen
  \bibfield  {author} {\bibinfo {author} {\bibfnamefont {P.~A.}\ \bibnamefont
  {McClarty}},\ }\bibfield  {title} {\bibinfo {title} {Topological magnons: A
  review},\ }\href {https://doi.org/10.1146/annurev-conmatphys-031620-104715}
  {\bibfield  {journal} {\bibinfo  {journal} {Annual Review of Condensed Matter
  Physics}\ }\textbf {\bibinfo {volume} {13}},\ \bibinfo {pages} {171}
  (\bibinfo {year} {2022})}\BibitemShut {NoStop}%
\bibitem [{\citenamefont {Zhang}\ and\ \citenamefont
  {Cheng}(2022)}]{2022_AFM_Nernst_Cheng_APL}%
  \BibitemOpen
  \bibfield  {author} {\bibinfo {author} {\bibfnamefont {H.}~\bibnamefont
  {Zhang}}\ and\ \bibinfo {author} {\bibfnamefont {R.}~\bibnamefont {Cheng}},\
  }\bibfield  {title} {\bibinfo {title} {{A perspective on magnon spin Nernst
  effect in antiferromagnets}},\ }\href {https://doi.org/10.1063/5.0084359}
  {\bibfield  {journal} {\bibinfo  {journal} {Applied Physics Letters}\
  }\textbf {\bibinfo {volume} {120}},\ \bibinfo {pages} {090502} (\bibinfo
  {year} {2022})}\BibitemShut {NoStop}%
\bibitem [{\citenamefont {Su}\ \emph {et~al.}(1979)\citenamefont {Su},
  \citenamefont {Schrieffer},\ and\ \citenamefont {Heeger}}]{1979_SSH_PRL}%
  \BibitemOpen
  \bibfield  {author} {\bibinfo {author} {\bibfnamefont {W.~P.}\ \bibnamefont
  {Su}}, \bibinfo {author} {\bibfnamefont {J.~R.}\ \bibnamefont {Schrieffer}},\
  and\ \bibinfo {author} {\bibfnamefont {A.~J.}\ \bibnamefont {Heeger}},\
  }\bibfield  {title} {\bibinfo {title} {Solitons in polyacetylene},\ }\href
  {https://doi.org/10.1103/PhysRevLett.42.1698} {\bibfield  {journal} {\bibinfo
   {journal} {Phys. Rev. Lett.}\ }\textbf {\bibinfo {volume} {42}},\ \bibinfo
  {pages} {1698} (\bibinfo {year} {1979})}\BibitemShut {NoStop}%
\bibitem [{\citenamefont {Pirmoradian}\ \emph {et~al.}(2018)\citenamefont
  {Pirmoradian}, \citenamefont {Zare~Rameshti}, \citenamefont {Miri},\ and\
  \citenamefont {Saeidian}}]{2018_SSH_Magnon_Chain}%
  \BibitemOpen
  \bibfield  {author} {\bibinfo {author} {\bibfnamefont {F.}~\bibnamefont
  {Pirmoradian}}, \bibinfo {author} {\bibfnamefont {B.}~\bibnamefont
  {Zare~Rameshti}}, \bibinfo {author} {\bibfnamefont {M.}~\bibnamefont
  {Miri}},\ and\ \bibinfo {author} {\bibfnamefont {S.}~\bibnamefont
  {Saeidian}},\ }\bibfield  {title} {\bibinfo {title} {Topological magnon modes
  in a chain of magnetic spheres},\ }\href
  {https://doi.org/10.1103/PhysRevB.98.224409} {\bibfield  {journal} {\bibinfo
  {journal} {Phys. Rev. B}\ }\textbf {\bibinfo {volume} {98}},\ \bibinfo
  {pages} {224409} (\bibinfo {year} {2018})}\BibitemShut {NoStop}%
\bibitem [{\citenamefont {Wei}\ \emph {et~al.}(2022)\citenamefont {Wei},
  \citenamefont {Ni}, \citenamefont {Zheng}, \citenamefont {Liu},\ and\
  \citenamefont {Zou}}]{2022_SSH_Magnons_JPCM}%
  \BibitemOpen
  \bibfield  {author} {\bibinfo {author} {\bibfnamefont {P.-T.}\ \bibnamefont
  {Wei}}, \bibinfo {author} {\bibfnamefont {J.-Y.}\ \bibnamefont {Ni}},
  \bibinfo {author} {\bibfnamefont {X.-M.}\ \bibnamefont {Zheng}}, \bibinfo
  {author} {\bibfnamefont {D.-Y.}\ \bibnamefont {Liu}},\ and\ \bibinfo {author}
  {\bibfnamefont {L.-J.}\ \bibnamefont {Zou}},\ }\bibfield  {title} {\bibinfo
  {title} {Topological magnons in one-dimensional ferromagnetic
  Su-Schrieffer-Heeger model with anisotropic interaction},\ }\href
  {https://doi.org/10.1088/1361-648X/ac99cb} {\bibfield  {journal} {\bibinfo
  {journal} {Journal of Physics: Condensed Matter}\ }\textbf {\bibinfo {volume}
  {34}},\ \bibinfo {pages} {495801} (\bibinfo {year} {2022})}\BibitemShut
  {NoStop}%
\bibitem [{\citenamefont {Str\"{o}bele}\ \emph {et~al.}(2016)\citenamefont
  {Str\"{o}bele}, \citenamefont {Thalwitzer},\ and\ \citenamefont
  {Meyer}}]{2016_MoI3_Facile_Synthesis}%
  \BibitemOpen
  \bibfield  {author} {\bibinfo {author} {\bibfnamefont {M.}~\bibnamefont
  {Str\"{o}bele}}, \bibinfo {author} {\bibfnamefont {R.}~\bibnamefont
  {Thalwitzer}},\ and\ \bibinfo {author} {\bibfnamefont {H.-J.}\ \bibnamefont
  {Meyer}},\ }\bibfield  {title} {\bibinfo {title} {Facile way of synthesis for
  molybdenum iodides},\ }\href {https://doi.org/10.1021/acs.inorgchem.6b02229}
  {\bibfield  {journal} {\bibinfo  {journal} {Inorganic Chemistry}\ }\textbf
  {\bibinfo {volume} {55}},\ \bibinfo {pages} {12074} (\bibinfo {year}
  {2016})}\BibitemShut {NoStop}%
\bibitem [{\citenamefont {Choi}\ \emph {et~al.}(2021)\citenamefont {Choi},
  \citenamefont {Oh}, \citenamefont {Chae}, \citenamefont {Jeong},
  \citenamefont {Kim}, \citenamefont {Jeon}, \citenamefont {Lee}, \citenamefont
  {Yoon}, \citenamefont {Woo}, \citenamefont {Dong}, \citenamefont {Ghulam},
  \citenamefont {Lim}, \citenamefont {Liu}, \citenamefont {Wang}, \citenamefont
  {Junaid}, \citenamefont {Lee}, \citenamefont {Yu},\ and\ \citenamefont
  {Choi}}]{MoI3_exp}%
  \BibitemOpen
  \bibfield  {author} {\bibinfo {author} {\bibfnamefont {K.~H.}\ \bibnamefont
  {Choi}}, \bibinfo {author} {\bibfnamefont {S.}~\bibnamefont {Oh}}, \bibinfo
  {author} {\bibfnamefont {S.}~\bibnamefont {Chae}}, \bibinfo {author}
  {\bibfnamefont {B.~J.}\ \bibnamefont {Jeong}}, \bibinfo {author}
  {\bibfnamefont {B.~J.}\ \bibnamefont {Kim}}, \bibinfo {author} {\bibfnamefont
  {J.}~\bibnamefont {Jeon}}, \bibinfo {author} {\bibfnamefont {S.~H.}\
  \bibnamefont {Lee}}, \bibinfo {author} {\bibfnamefont {S.~O.}\ \bibnamefont
  {Yoon}}, \bibinfo {author} {\bibfnamefont {C.}~\bibnamefont {Woo}}, \bibinfo
  {author} {\bibfnamefont {X.}~\bibnamefont {Dong}}, \bibinfo {author}
  {\bibfnamefont {A.}~\bibnamefont {Ghulam}}, \bibinfo {author} {\bibfnamefont
  {C.}~\bibnamefont {Lim}}, \bibinfo {author} {\bibfnamefont {Z.}~\bibnamefont
  {Liu}}, \bibinfo {author} {\bibfnamefont {C.}~\bibnamefont {Wang}}, \bibinfo
  {author} {\bibfnamefont {A.}~\bibnamefont {Junaid}}, \bibinfo {author}
  {\bibfnamefont {J.-H.}\ \bibnamefont {Lee}}, \bibinfo {author} {\bibfnamefont
  {H.~K.}\ \bibnamefont {Yu}},\ and\ \bibinfo {author} {\bibfnamefont {J.-Y.}\
  \bibnamefont {Choi}},\ }\bibfield  {title} {\bibinfo {title} {Low ligand
  field strength ion {(I-)} mediated {1D} inorganic material {MoI$_3$}:
  Synthesis and application to photo-detectors},\ }\href
  {https://doi.org/10.1016/j.jallcom.2020.157375} {\bibfield  {journal}
  {\bibinfo  {journal} {Journal of Alloys and Compounds}\ }\textbf {\bibinfo
  {volume} {853}},\ \bibinfo {pages} {157375} (\bibinfo {year}
  {2021})}\BibitemShut {NoStop}%
\bibitem [{\citenamefont {Kargar}\ \emph {et~al.}(2022)\citenamefont {Kargar},
  \citenamefont {Barani}, \citenamefont {Sesing}, \citenamefont {Mai},
  \citenamefont {Debnath}, \citenamefont {Zhang}, \citenamefont {Liu},
  \citenamefont {Zhu}, \citenamefont {Ghosh}, \citenamefont {Biacchi},
  \citenamefont {da~Jornada}, \citenamefont {Bartels}, \citenamefont {Adel},
  \citenamefont {Hight~Walker}, \citenamefont {Davydov}, \citenamefont
  {Salguero}, \citenamefont {Lake},\ and\ \citenamefont
  {Balandin}}]{2022_MoI3_Fari_APL}%
  \BibitemOpen
  \bibfield  {author} {\bibinfo {author} {\bibfnamefont {F.}~\bibnamefont
  {Kargar}}, \bibinfo {author} {\bibfnamefont {Z.}~\bibnamefont {Barani}},
  \bibinfo {author} {\bibfnamefont {N.~R.}\ \bibnamefont {Sesing}}, \bibinfo
  {author} {\bibfnamefont {T.~T.}\ \bibnamefont {Mai}}, \bibinfo {author}
  {\bibfnamefont {T.}~\bibnamefont {Debnath}}, \bibinfo {author} {\bibfnamefont
  {H.}~\bibnamefont {Zhang}}, \bibinfo {author} {\bibfnamefont
  {Y.}~\bibnamefont {Liu}}, \bibinfo {author} {\bibfnamefont {Y.}~\bibnamefont
  {Zhu}}, \bibinfo {author} {\bibfnamefont {S.}~\bibnamefont {Ghosh}}, \bibinfo
  {author} {\bibfnamefont {A.~J.}\ \bibnamefont {Biacchi}}, \bibinfo {author}
  {\bibfnamefont {F.~H.}\ \bibnamefont {da~Jornada}}, \bibinfo {author}
  {\bibfnamefont {L.}~\bibnamefont {Bartels}}, \bibinfo {author} {\bibfnamefont
  {T.}~\bibnamefont {Adel}}, \bibinfo {author} {\bibfnamefont {A.~R.}\
  \bibnamefont {Hight~Walker}}, \bibinfo {author} {\bibfnamefont {A.~V.}\
  \bibnamefont {Davydov}}, \bibinfo {author} {\bibfnamefont {T.~T.}\
  \bibnamefont {Salguero}}, \bibinfo {author} {\bibfnamefont {R.~K.}\
  \bibnamefont {Lake}},\ and\ \bibinfo {author} {\bibfnamefont {A.~A.}\
  \bibnamefont {Balandin}},\ }\bibfield  {title} {\bibinfo {title} {Elemental
  excitations in {MoI$_3$} one-dimensional van der waals nanowires},\ }\href
  {https://doi.org/10.1063/5.0129904} {\bibfield  {journal} {\bibinfo
  {journal} {Applied Physics Letters}\ }\textbf {\bibinfo {volume} {121}},\
  \bibinfo {pages} {221901} (\bibinfo {year} {2022})}\BibitemShut {NoStop}%
\bibitem [{\citenamefont {Fu}\ \emph {et~al.}(2022)\citenamefont {Fu},
  \citenamefont {Shang}, \citenamefont {Zhou}, \citenamefont {Guo},\ and\
  \citenamefont {Zhao}}]{2022_TM_Halide_Wires_APL}%
  \BibitemOpen
  \bibfield  {author} {\bibinfo {author} {\bibfnamefont {L.}~\bibnamefont
  {Fu}}, \bibinfo {author} {\bibfnamefont {C.}~\bibnamefont {Shang}}, \bibinfo
  {author} {\bibfnamefont {S.}~\bibnamefont {Zhou}}, \bibinfo {author}
  {\bibfnamefont {Y.}~\bibnamefont {Guo}},\ and\ \bibinfo {author}
  {\bibfnamefont {J.}~\bibnamefont {Zhao}},\ }\bibfield  {title} {\bibinfo
  {title} {{Transition metal halide nanowires: A family of one-dimensional
  multifunctional building blocks}},\ }\href
  {https://doi.org/10.1063/5.0078819} {\bibfield  {journal} {\bibinfo
  {journal} {Applied Physics Letters}\ }\textbf {\bibinfo {volume} {120}},\
  \bibinfo {pages} {023103} (\bibinfo {year} {2022})}\BibitemShut {NoStop}%
\bibitem [{\citenamefont {Mella}\ \emph {et~al.}(2024)\citenamefont {Mella},
  \citenamefont {Suárez-Morell},\ and\ \citenamefont
  {Nunez}}]{2024_MoI3_Spin_Chains_JMMM}%
  \BibitemOpen
  \bibfield  {author} {\bibinfo {author} {\bibfnamefont {A.}~\bibnamefont
  {Mella}}, \bibinfo {author} {\bibfnamefont {E.}~\bibnamefont
  {Suárez-Morell}},\ and\ \bibinfo {author} {\bibfnamefont {A.~S.}\
  \bibnamefont {Nunez}},\ }\bibfield  {title} {\bibinfo {title} {Magnetic
  spirals and biquadratic exchange in {1D MoX$_3$} spin chains},\ }\href
  {https://doi.org/10.1016/j.jmmm.2024.171882} {\bibfield  {journal} {\bibinfo
  {journal} {Journal of Magnetism and Magnetic Materials}\ }\textbf {\bibinfo
  {volume} {594}},\ \bibinfo {pages} {171882} (\bibinfo {year}
  {2024})}\BibitemShut {NoStop}%
\bibitem [{\citenamefont {Momma}\ and\ \citenamefont {Izumi}(2011)}]{VESTA}%
  \BibitemOpen
  \bibfield  {author} {\bibinfo {author} {\bibfnamefont {K.}~\bibnamefont
  {Momma}}\ and\ \bibinfo {author} {\bibfnamefont {F.}~\bibnamefont {Izumi}},\
  }\bibfield  {title} {\bibinfo {title} {{{\it VESTA3} for three-dimensional
  visualization of crystal, volumetric and morphology data}},\ }\href
  {https://doi.org/10.1107/S0021889811038970} {\bibfield  {journal} {\bibinfo
  {journal} {Journal of Applied Crystallography}\ }\textbf {\bibinfo {volume}
  {44}},\ \bibinfo {pages} {1272} (\bibinfo {year} {2011})}\BibitemShut
  {NoStop}%
\bibitem [{\citenamefont {Toth}\ and\ \citenamefont
  {Lake}(2015)}]{SpinW_Toth_2015}%
  \BibitemOpen
  \bibfield  {author} {\bibinfo {author} {\bibfnamefont {S.}~\bibnamefont
  {Toth}}\ and\ \bibinfo {author} {\bibfnamefont {B.}~\bibnamefont {Lake}},\
  }\bibfield  {title} {\bibinfo {title} {Linear spin wave theory for {single-Q}
  incommensurate magnetic structures},\ }\href
  {https://doi.org/10.1088/0953-8984/27/16/166002} {\bibfield  {journal}
  {\bibinfo  {journal} {Journal of Physics: Condensed Matter}\ }\textbf
  {\bibinfo {volume} {27}},\ \bibinfo {pages} {166002} (\bibinfo {year}
  {2015})}\BibitemShut {NoStop}%
\bibitem [{\citenamefont {Yosida}(1991)}]{Yosida_Th_Mag}%
  \BibitemOpen
  \bibfield  {author} {\bibinfo {author} {\bibfnamefont {K.}~\bibnamefont
  {Yosida}},\ }\href@noop {} {\emph {\bibinfo {title} {Theory of Magnetism}}}\
  (\bibinfo  {publisher} {Springer-Verlag},\ \bibinfo {address} {New York},\
  \bibinfo {year} {1991})\BibitemShut {NoStop}%
\bibitem [{\citenamefont {Nolting}\ and\ \citenamefont
  {Ramakanth}(2009)}]{Nolting_Qu_Mag}%
  \BibitemOpen
  \bibfield  {author} {\bibinfo {author} {\bibfnamefont {W.}~\bibnamefont
  {Nolting}}\ and\ \bibinfo {author} {\bibfnamefont {A.}~\bibnamefont
  {Ramakanth}},\ }\href {https://doi.org/10.1007/978-3-540-85416-6} {\emph
  {\bibinfo {title} {Quantum Theory of Magnetism}}}\ (\bibinfo  {publisher}
  {Springer},\ \bibinfo {address} {New York},\ \bibinfo {year}
  {2009})\BibitemShut {NoStop}%
\bibitem [{\citenamefont {Rezende}(2020)}]{2020_Rezende_Fund_Magnonics}%
  \BibitemOpen
  \bibfield  {author} {\bibinfo {author} {\bibfnamefont {S.~M.}\ \bibnamefont
  {Rezende}},\ }\href@noop {} {\emph {\bibinfo {title} {Lecture Notes in
  Physics - Fundamentals of Magnonics}}},\ Vol.\ \bibinfo {volume} {969}\
  (\bibinfo  {publisher} {Springer Nature Switzerland AG},\ \bibinfo {address}
  {Cham},\ \bibinfo {year} {2020})\BibitemShut {NoStop}%
\bibitem [{\citenamefont {Kresse}\ and\ \citenamefont
  {Hafner}(1993)}]{VASP:I:PRB:1993}%
  \BibitemOpen
  \bibfield  {author} {\bibinfo {author} {\bibfnamefont {G.}~\bibnamefont
  {Kresse}}\ and\ \bibinfo {author} {\bibfnamefont {J.}~\bibnamefont
  {Hafner}},\ }\bibfield  {title} {\bibinfo {title} {Ab initio molecular
  dynamics for liquid metals},\ }\href
  {https://doi.org/10.1103/PhysRevB.47.558} {\bibfield  {journal} {\bibinfo
  {journal} {Phys. Rev. B}\ }\textbf {\bibinfo {volume} {47}},\ \bibinfo
  {pages} {558} (\bibinfo {year} {1993})}\BibitemShut {NoStop}%
\bibitem [{\citenamefont {Kresse}\ and\ \citenamefont
  {Hafner}(1994)}]{VASP:II:PRB:1994}%
  \BibitemOpen
  \bibfield  {author} {\bibinfo {author} {\bibfnamefont {G.}~\bibnamefont
  {Kresse}}\ and\ \bibinfo {author} {\bibfnamefont {J.}~\bibnamefont
  {Hafner}},\ }\bibfield  {title} {\bibinfo {title} {Ab initio
  molecular-dynamics simulation of the liquid-metal--amorphous-semiconductor
  transition in germanium},\ }\href {https://doi.org/10.1103/PhysRevB.49.14251}
  {\bibfield  {journal} {\bibinfo  {journal} {Phys. Rev. B}\ }\textbf {\bibinfo
  {volume} {49}},\ \bibinfo {pages} {14251} (\bibinfo {year}
  {1994})}\BibitemShut {NoStop}%
\bibitem [{\citenamefont {Kresse}\ and\ \citenamefont
  {Furthm\"uller}(1996)}]{VASP:III:PRB:1996}%
  \BibitemOpen
  \bibfield  {author} {\bibinfo {author} {\bibfnamefont {G.}~\bibnamefont
  {Kresse}}\ and\ \bibinfo {author} {\bibfnamefont {J.}~\bibnamefont
  {Furthm\"uller}},\ }\bibfield  {title} {\bibinfo {title} {Efficient iterative
  schemes for ab initio total-energy calculations using a plane-wave basis
  set},\ }\href {https://doi.org/10.1103/PhysRevB.54.11169} {\bibfield
  {journal} {\bibinfo  {journal} {Phys. Rev. B}\ }\textbf {\bibinfo {volume}
  {54}},\ \bibinfo {pages} {11169} (\bibinfo {year} {1996})}\BibitemShut
  {NoStop}%
\bibitem [{\citenamefont {Kresse}\ and\ \citenamefont
  {Furthmuller}(1996)}]{VASP4_CompMatSci}%
  \BibitemOpen
  \bibfield  {author} {\bibinfo {author} {\bibfnamefont {G.}~\bibnamefont
  {Kresse}}\ and\ \bibinfo {author} {\bibfnamefont {J.}~\bibnamefont
  {Furthmuller}},\ }\bibfield  {title} {\bibinfo {title} {Efficiency of
  ab-initio total energy calculations for metals and semiconductors using a
  plane-wave basis set},\ }\href@noop {} {\bibfield  {journal} {\bibinfo
  {journal} {Comput. Mater. Sci.}\ }\textbf {\bibinfo {volume} {6}},\ \bibinfo
  {pages} {15} (\bibinfo {year} {1996})}\BibitemShut {NoStop}%
\bibitem [{\citenamefont {Perdew}\ \emph {et~al.}(1996)\citenamefont {Perdew},
  \citenamefont {Burke},\ and\ \citenamefont {Ernzerhof}}]{PBE}%
  \BibitemOpen
  \bibfield  {author} {\bibinfo {author} {\bibfnamefont {J.~P.}\ \bibnamefont
  {Perdew}}, \bibinfo {author} {\bibfnamefont {K.}~\bibnamefont {Burke}},\ and\
  \bibinfo {author} {\bibfnamefont {M.}~\bibnamefont {Ernzerhof}},\ }\bibfield
  {title} {\bibinfo {title} {Generalized gradient approximation made simple},\
  }\href {https://doi.org/10.1103/PhysRevLett.77.3865} {\bibfield  {journal}
  {\bibinfo  {journal} {Phys. Rev. Lett.}\ }\textbf {\bibinfo {volume} {77}},\
  \bibinfo {pages} {3865} (\bibinfo {year} {1996})}\BibitemShut {NoStop}%
\bibitem [{\citenamefont {Grimme}\ \emph {et~al.}(2010)\citenamefont {Grimme},
  \citenamefont {Antony}, \citenamefont {Ehrlich},\ and\ \citenamefont
  {Krieg}}]{grimme2010consistent}%
  \BibitemOpen
  \bibfield  {author} {\bibinfo {author} {\bibfnamefont {S.}~\bibnamefont
  {Grimme}}, \bibinfo {author} {\bibfnamefont {J.}~\bibnamefont {Antony}},
  \bibinfo {author} {\bibfnamefont {S.}~\bibnamefont {Ehrlich}},\ and\ \bibinfo
  {author} {\bibfnamefont {H.}~\bibnamefont {Krieg}},\ }\bibfield  {title}
  {\bibinfo {title} {A consistent and accurate ab initio parametrization of
  density functional dispersion correction {(DFT-D)} for the 94 elements
  {H-Pu}},\ }\href {https://doi.org/10.1063/1.3382344} {\bibfield  {journal}
  {\bibinfo  {journal} {The Journal of Chemical Physics}\ }\textbf {\bibinfo
  {volume} {132}},\ \bibinfo {pages} {154104} (\bibinfo {year}
  {2010})}\BibitemShut {NoStop}%
\bibitem [{\citenamefont {Togo}\ \emph {et~al.}(2023)\citenamefont {Togo},
  \citenamefont {Chaput}, \citenamefont {Tadano},\ and\ \citenamefont
  {Tanaka}}]{phonopy-phono3py-JPCM}%
  \BibitemOpen
  \bibfield  {author} {\bibinfo {author} {\bibfnamefont {A.}~\bibnamefont
  {Togo}}, \bibinfo {author} {\bibfnamefont {L.}~\bibnamefont {Chaput}},
  \bibinfo {author} {\bibfnamefont {T.}~\bibnamefont {Tadano}},\ and\ \bibinfo
  {author} {\bibfnamefont {I.}~\bibnamefont {Tanaka}},\ }\bibfield  {title}
  {\bibinfo {title} {Implementation strategies in phonopy and phono3py},\
  }\href {https://doi.org/10.1088/1361-648X/acd831} {\bibfield  {journal}
  {\bibinfo  {journal} {J. Phys. Condens. Matter}\ }\textbf {\bibinfo {volume}
  {35}},\ \bibinfo {pages} {353001} (\bibinfo {year} {2023})}\BibitemShut
  {NoStop}%
\bibitem [{\citenamefont {Togo}(2023)}]{phonopy-phono3py-JPSJ}%
  \BibitemOpen
  \bibfield  {author} {\bibinfo {author} {\bibfnamefont {A.}~\bibnamefont
  {Togo}},\ }\bibfield  {title} {\bibinfo {title} {First-principles phonon
  calculations with phonopy and phono3py},\ }\href
  {https://doi.org/10.7566/JPSJ.92.012001} {\bibfield  {journal} {\bibinfo
  {journal} {J. Phys. Soc. Jpn.}\ }\textbf {\bibinfo {volume} {92}},\ \bibinfo
  {pages} {012001} (\bibinfo {year} {2023})}\BibitemShut {NoStop}%
\bibitem [{\citenamefont {Calderon}\ \emph {et~al.}(2015)\citenamefont
  {Calderon}, \citenamefont {Plata}, \citenamefont {Toher}, \citenamefont
  {Oses}, \citenamefont {Levy}, \citenamefont {Fornari}, \citenamefont {Natan},
  \citenamefont {Mehl}, \citenamefont {Hart}, \citenamefont {{Buongiorno
  Nardelli}},\ and\ \citenamefont
  {Curtarolo}}]{AFLOW_Hi_Thruput_CompMatSci_2015}%
  \BibitemOpen
  \bibfield  {author} {\bibinfo {author} {\bibfnamefont {C.~E.}\ \bibnamefont
  {Calderon}}, \bibinfo {author} {\bibfnamefont {J.~J.}\ \bibnamefont {Plata}},
  \bibinfo {author} {\bibfnamefont {C.}~\bibnamefont {Toher}}, \bibinfo
  {author} {\bibfnamefont {C.}~\bibnamefont {Oses}}, \bibinfo {author}
  {\bibfnamefont {O.}~\bibnamefont {Levy}}, \bibinfo {author} {\bibfnamefont
  {M.}~\bibnamefont {Fornari}}, \bibinfo {author} {\bibfnamefont
  {A.}~\bibnamefont {Natan}}, \bibinfo {author} {\bibfnamefont {M.~J.}\
  \bibnamefont {Mehl}}, \bibinfo {author} {\bibfnamefont {G.}~\bibnamefont
  {Hart}}, \bibinfo {author} {\bibfnamefont {M.}~\bibnamefont {{Buongiorno
  Nardelli}}},\ and\ \bibinfo {author} {\bibfnamefont {S.}~\bibnamefont
  {Curtarolo}},\ }\bibfield  {title} {\bibinfo {title} {The {AFLOW} standard
  for high-throughput materials science calculations},\ }\href
  {https://doi.org/https://doi.org/10.1016/j.commatsci.2015.07.019} {\bibfield
  {journal} {\bibinfo  {journal} {Computational Materials Science}\ }\textbf
  {\bibinfo {volume} {108}},\ \bibinfo {pages} {233} (\bibinfo {year}
  {2015})}\BibitemShut {NoStop}%
\bibitem [{\citenamefont {Kang}\ \emph {et~al.}(2009)\citenamefont {Kang},
  \citenamefont {Lee}, \citenamefont {Kremer},\ and\ \citenamefont
  {Whangbo}}]{2009_Whangbo_diamon_chain_JPCM}%
  \BibitemOpen
  \bibfield  {author} {\bibinfo {author} {\bibfnamefont {J.}~\bibnamefont
  {Kang}}, \bibinfo {author} {\bibfnamefont {C.}~\bibnamefont {Lee}}, \bibinfo
  {author} {\bibfnamefont {R.~K.}\ \bibnamefont {Kremer}},\ and\ \bibinfo
  {author} {\bibfnamefont {M.-H.}\ \bibnamefont {Whangbo}},\ }\bibfield
  {title} {\bibinfo {title} {Consequences of the intrachain dimer-monomer spin
  frustration and the interchain dimer-monomer spin exchange in the
  diamond-chain compound azurite {Cu$_3$(CO$_3$)$_2$(OH)$_2$}},\ }\href
  {https://doi.org/10.1088/0953-8984/21/39/392201} {\bibfield  {journal}
  {\bibinfo  {journal} {Journal of Physics: Condensed Matter}\ }\textbf
  {\bibinfo {volume} {21}},\ \bibinfo {pages} {392201} (\bibinfo {year}
  {2009})}\BibitemShut {NoStop}%
\bibitem [{\citenamefont {Lin}\ \emph {et~al.}(2022)\citenamefont {Lin},
  \citenamefont {Soni}, \citenamefont {Zhang}, \citenamefont {Gao},
  \citenamefont {Moreo}, \citenamefont {Alvarez}, \citenamefont {Christianson},
  \citenamefont {Stone},\ and\ \citenamefont
  {Dagotto}}]{2022_Na2Cu2TeO6_Dagatto_PRB}%
  \BibitemOpen
  \bibfield  {author} {\bibinfo {author} {\bibfnamefont {L.-F.}\ \bibnamefont
  {Lin}}, \bibinfo {author} {\bibfnamefont {R.}~\bibnamefont {Soni}}, \bibinfo
  {author} {\bibfnamefont {Y.}~\bibnamefont {Zhang}}, \bibinfo {author}
  {\bibfnamefont {S.}~\bibnamefont {Gao}}, \bibinfo {author} {\bibfnamefont
  {A.}~\bibnamefont {Moreo}}, \bibinfo {author} {\bibfnamefont
  {G.}~\bibnamefont {Alvarez}}, \bibinfo {author} {\bibfnamefont {A.~D.}\
  \bibnamefont {Christianson}}, \bibinfo {author} {\bibfnamefont {M.~B.}\
  \bibnamefont {Stone}},\ and\ \bibinfo {author} {\bibfnamefont
  {E.}~\bibnamefont {Dagotto}},\ }\bibfield  {title} {\bibinfo {title}
  {Electronic structure, magnetic properties, and pairing tendencies of the
  copper-based honeycomb lattice
  {${\mathrm{Na}}_{2}{\mathrm{Cu}}_{2}{\mathrm{TeO}}_{6}$}},\ }\href
  {https://doi.org/10.1103/PhysRevB.105.245113} {\bibfield  {journal} {\bibinfo
   {journal} {Phys. Rev. B}\ }\textbf {\bibinfo {volume} {105}},\ \bibinfo
  {pages} {245113} (\bibinfo {year} {2022})}\BibitemShut {NoStop}%
\bibitem [{\citenamefont {Liu}\ \emph {et~al.}(2022)\citenamefont {Liu},
  \citenamefont {Kwon}, \citenamefont {de~Coster}, \citenamefont {Lake},\ and\
  \citenamefont {Neupane}}]{2022_CrTe2_YLiu_PRMat}%
  \BibitemOpen
  \bibfield  {author} {\bibinfo {author} {\bibfnamefont {Y.}~\bibnamefont
  {Liu}}, \bibinfo {author} {\bibfnamefont {S.}~\bibnamefont {Kwon}}, \bibinfo
  {author} {\bibfnamefont {G.~J.}\ \bibnamefont {de~Coster}}, \bibinfo {author}
  {\bibfnamefont {R.~K.}\ \bibnamefont {Lake}},\ and\ \bibinfo {author}
  {\bibfnamefont {M.~R.}\ \bibnamefont {Neupane}},\ }\bibfield  {title}
  {\bibinfo {title} {Structural, electronic, and magnetic properties of
  {CrTe$_2$}},\ }\href {https://doi.org/10.1103/PhysRevMaterials.6.084004}
  {\bibfield  {journal} {\bibinfo  {journal} {Phys. Rev. Mater.}\ }\textbf
  {\bibinfo {volume} {6}},\ \bibinfo {pages} {084004} (\bibinfo {year}
  {2022})}\BibitemShut {NoStop}%
\bibitem [{\citenamefont {Sancho}\ \emph {et~al.}(1984)\citenamefont {Sancho},
  \citenamefont {Sancho},\ and\ \citenamefont {Rubio}}]{Sancho_Rubio_JPhysF84}%
  \BibitemOpen
  \bibfield  {author} {\bibinfo {author} {\bibfnamefont {M.~P.~L.}\
  \bibnamefont {Sancho}}, \bibinfo {author} {\bibfnamefont {J.~M.~L.}\
  \bibnamefont {Sancho}},\ and\ \bibinfo {author} {\bibfnamefont
  {J.}~\bibnamefont {Rubio}},\ }\bibfield  {title} {\bibinfo {title} {Quick
  iterative scheme for the calculation of transfer matrices: application to
  {Mo} (100)},\ }\href {http://dx.doi.org/10.1088/0305-4608/14/5/016}
  {\bibfield  {journal} {\bibinfo  {journal} {Journal of Physics F: Metal
  Physics}\ }\textbf {\bibinfo {volume} {14}},\ \bibinfo {pages} {1205}
  (\bibinfo {year} {1984})}\BibitemShut {NoStop}%
\bibitem [{\citenamefont {Sancho}\ \emph {et~al.}(1985)\citenamefont {Sancho},
  \citenamefont {Sancho},\ and\ \citenamefont
  {Rubio}}]{MPLSancho_HighlyConvergent_JPF85}%
  \BibitemOpen
  \bibfield  {author} {\bibinfo {author} {\bibfnamefont {M.~P.~L.}\
  \bibnamefont {Sancho}}, \bibinfo {author} {\bibfnamefont {J.~M.~L.}\
  \bibnamefont {Sancho}},\ and\ \bibinfo {author} {\bibfnamefont
  {J.}~\bibnamefont {Rubio}},\ }\bibfield  {title} {\bibinfo {title} {Highly
  convergent schemes for the calculation of bulk and surface {Green}
  functions},\ }\href@noop {} {\bibfield  {journal} {\bibinfo  {journal} {J.
  Phys. F}\ }\textbf {\bibinfo {volume} {15}},\ \bibinfo {pages} {851}
  (\bibinfo {year} {1985})}\BibitemShut {NoStop}%
\bibitem [{\citenamefont {Galperin}\ \emph {et~al.}(2002)\citenamefont
  {Galperin}, \citenamefont {Toledo},\ and\ \citenamefont
  {Nitzan}}]{Galperin:JCP:2002:decimation}%
  \BibitemOpen
  \bibfield  {author} {\bibinfo {author} {\bibfnamefont {M.}~\bibnamefont
  {Galperin}}, \bibinfo {author} {\bibfnamefont {S.}~\bibnamefont {Toledo}},\
  and\ \bibinfo {author} {\bibfnamefont {A.}~\bibnamefont {Nitzan}},\
  }\bibfield  {title} {\bibinfo {title} {Numerical computation of tunneling
  fluxes},\ }\href@noop {} {\bibfield  {journal} {\bibinfo  {journal} {J. Chem.
  Phys.}\ }\textbf {\bibinfo {volume} {117}},\ \bibinfo {pages} {10817}
  (\bibinfo {year} {2002})}\BibitemShut {NoStop}%
\bibitem [{\citenamefont {Holstein}\ and\ \citenamefont
  {Primakoff}(1940)}]{Holstein_Primakoff1940}%
  \BibitemOpen
  \bibfield  {author} {\bibinfo {author} {\bibfnamefont {T.}~\bibnamefont
  {Holstein}}\ and\ \bibinfo {author} {\bibfnamefont {H.}~\bibnamefont
  {Primakoff}},\ }\bibfield  {title} {\bibinfo {title} {Field dependence of the
  intrinsic domain magnetization of a ferromagnet},\ }\href@noop {} {\bibfield
  {journal} {\bibinfo  {journal} {Phys. Rev.}\ }\textbf {\bibinfo {volume}
  {58}},\ \bibinfo {pages} {1098} (\bibinfo {year} {1940})}\BibitemShut
  {NoStop}%
\bibitem [{\citenamefont {White}\ \emph {et~al.}(1965)\citenamefont {White},
  \citenamefont {Sparks},\ and\ \citenamefont
  {Ortenburger}}]{1965_Diag_AFM_Magnon_Phonon_Interaction}%
  \BibitemOpen
  \bibfield  {author} {\bibinfo {author} {\bibfnamefont {R.~M.}\ \bibnamefont
  {White}}, \bibinfo {author} {\bibfnamefont {M.}~\bibnamefont {Sparks}},\ and\
  \bibinfo {author} {\bibfnamefont {I.}~\bibnamefont {Ortenburger}},\
  }\bibfield  {title} {\bibinfo {title} {Diagonalization of the
  antiferromagnetic magnon-phonon interaction},\ }\href
  {https://doi.org/10.1103/PhysRev.139.A450} {\bibfield  {journal} {\bibinfo
  {journal} {Phys. Rev.}\ }\textbf {\bibinfo {volume} {139}},\ \bibinfo {pages}
  {A450} (\bibinfo {year} {1965})}\BibitemShut {NoStop}%
\bibitem [{\citenamefont {Colpa}(1978)}]{1978_Colpa}%
  \BibitemOpen
  \bibfield  {author} {\bibinfo {author} {\bibfnamefont {J.~H.~P.}\
  \bibnamefont {Colpa}},\ }\bibfield  {title} {\bibinfo {title}
  {Diagonalization of the quadratic boson {Hamiltonian}},\ }\href@noop {}
  {\bibfield  {journal} {\bibinfo  {journal} {Physica A: Statistical Mechanics
  and its Applications}\ }\textbf {\bibinfo {volume} {93}},\ \bibinfo {pages}
  {327} (\bibinfo {year} {1978})}\BibitemShut {NoStop}%
\bibitem [{\citenamefont {Kittel}(1996)}]{Kittel}%
  \BibitemOpen
  \bibfield  {author} {\bibinfo {author} {\bibfnamefont {C.}~\bibnamefont
  {Kittel}},\ }\href@noop {} {\emph {\bibinfo {title} {Introduction to Solid
  State Physics}}},\ \bibinfo {edition} {7th}\ ed.\ (\bibinfo  {publisher}
  {John Wiley \& Sons},\ \bibinfo {address} {New York},\ \bibinfo {year}
  {1996})\BibitemShut {NoStop}%
\bibitem [{\citenamefont {Shockley}(1939)}]{1939_Surface_States_Shockley}%
  \BibitemOpen
  \bibfield  {author} {\bibinfo {author} {\bibfnamefont {W.}~\bibnamefont
  {Shockley}},\ }\bibfield  {title} {\bibinfo {title} {On the surface states
  associated with a periodic potential},\ }\href@noop {} {\bibfield  {journal}
  {\bibinfo  {journal} {Physical Review}\ }\textbf {\bibinfo {volume} {56}},\
  \bibinfo {pages} {317} (\bibinfo {year} {1939})}\BibitemShut {NoStop}%
\bibitem [{\citenamefont {Boerner}\ \emph {et~al.}(2023)\citenamefont
  {Boerner}, \citenamefont {Deems}, \citenamefont {Furlani}, \citenamefont
  {Knuth},\ and\ \citenamefont {Towns}}]{ACCESS}%
  \BibitemOpen
  \bibfield  {author} {\bibinfo {author} {\bibfnamefont {T.~J.}\ \bibnamefont
  {Boerner}}, \bibinfo {author} {\bibfnamefont {S.}~\bibnamefont {Deems}},
  \bibinfo {author} {\bibfnamefont {T.~R.}\ \bibnamefont {Furlani}}, \bibinfo
  {author} {\bibfnamefont {S.~L.}\ \bibnamefont {Knuth}},\ and\ \bibinfo
  {author} {\bibfnamefont {J.}~\bibnamefont {Towns}},\ }\bibfield  {title}
  {\bibinfo {title} {Access: Advancing innovation: Nsf's advanced
  cyberinfrastructure coordination ecosystem: Services \& support},\ }in\ \href
  {https://doi.org/10.1145/3569951.3597559} {\emph {\bibinfo {booktitle}
  {Practice and Experience in Advanced Research Computing}}},\ \bibinfo {series
  and number} {PEARC '23}\ (\bibinfo  {publisher} {Association for Computing
  Machinery},\ \bibinfo {address} {New York, NY, USA},\ \bibinfo {year}
  {2023})\ pp.\ \bibinfo {pages} {173--176}\BibitemShut {NoStop}%
\bibitem [{\citenamefont {\ifmmode~\check{S}\else \v{S}\fi{}abani}\ \emph
  {et~al.}(2020)\citenamefont {\ifmmode~\check{S}\else \v{S}\fi{}abani},
  \citenamefont {Bacaksiz},\ and\ \citenamefont {Milo\ifmmode \check{s}\else
  \v{s}\fi{}evi\ifmmode~\acute{c}\else
  \'{c}\fi{}}}]{2020_Generic_4_State_E_Map_Jij_PRB}%
  \BibitemOpen
  \bibfield  {author} {\bibinfo {author} {\bibfnamefont {D.}~\bibnamefont
  {\ifmmode~\check{S}\else \v{S}\fi{}abani}}, \bibinfo {author} {\bibfnamefont
  {C.}~\bibnamefont {Bacaksiz}},\ and\ \bibinfo {author} {\bibfnamefont
  {M.~V.}\ \bibnamefont {Milo\ifmmode \check{s}\else
  \v{s}\fi{}evi\ifmmode~\acute{c}\else \'{c}\fi{}}},\ }\bibfield  {title}
  {\bibinfo {title} {Ab initio methodology for magnetic exchange parameters:
  Generic four-state energy mapping onto a {Heisenberg} spin hamiltonian},\
  }\href {https://doi.org/10.1103/PhysRevB.102.014457} {\bibfield  {journal}
  {\bibinfo  {journal} {Phys. Rev. B}\ }\textbf {\bibinfo {volume} {102}},\
  \bibinfo {pages} {014457} (\bibinfo {year} {2020})}\BibitemShut {NoStop}%
\bibitem [{\citenamefont {Tiwari}\ \emph {et~al.}(2021)\citenamefont {Tiwari},
  \citenamefont {Van~de Put}, \citenamefont {Sor\'ee},\ and\ \citenamefont
  {Vandenberghe}}]{2021_2D_FMs_Vandenberghe_PRB}%
  \BibitemOpen
  \bibfield  {author} {\bibinfo {author} {\bibfnamefont {S.}~\bibnamefont
  {Tiwari}}, \bibinfo {author} {\bibfnamefont {M.~L.}\ \bibnamefont {Van~de
  Put}}, \bibinfo {author} {\bibfnamefont {B.}~\bibnamefont {Sor\'ee}},\ and\
  \bibinfo {author} {\bibfnamefont {W.~G.}\ \bibnamefont {Vandenberghe}},\
  }\bibfield  {title} {\bibinfo {title} {Critical behavior of the ferromagnets
  {${\mathrm{CrI}}_{3}, {\mathrm{CrBr}}_{3}$}, and {${\mathrm{CrGeTe}}_{3}$}
  and the antiferromagnet {${\mathrm{FeCl}}_{2}$}: A detailed first-principles
  study},\ }\href {https://doi.org/10.1103/PhysRevB.103.014432} {\bibfield
  {journal} {\bibinfo  {journal} {Phys. Rev. B}\ }\textbf {\bibinfo {volume}
  {103}},\ \bibinfo {pages} {014432} (\bibinfo {year} {2021})}\BibitemShut
  {NoStop}%
\bibitem [{\citenamefont {Korotin}\ \emph {et~al.}(2015)\citenamefont
  {Korotin}, \citenamefont {Mazurenko}, \citenamefont {Anisimov},\ and\
  \citenamefont {Streltsov}}]{2015_Green_Fn_Calc_J_PRB}%
  \BibitemOpen
  \bibfield  {author} {\bibinfo {author} {\bibfnamefont {D.~M.}\ \bibnamefont
  {Korotin}}, \bibinfo {author} {\bibfnamefont {V.~V.}\ \bibnamefont
  {Mazurenko}}, \bibinfo {author} {\bibfnamefont {V.~I.}\ \bibnamefont
  {Anisimov}},\ and\ \bibinfo {author} {\bibfnamefont {S.~V.}\ \bibnamefont
  {Streltsov}},\ }\bibfield  {title} {\bibinfo {title} {Calculation of exchange
  constants of the heisenberg model in plane-wave-based methods using the
  Green's function approach},\ }\href
  {https://doi.org/10.1103/PhysRevB.91.224405} {\bibfield  {journal} {\bibinfo
  {journal} {Phys. Rev. B}\ }\textbf {\bibinfo {volume} {91}},\ \bibinfo
  {pages} {224405} (\bibinfo {year} {2015})}\BibitemShut {NoStop}%
\bibitem [{\citenamefont {Terasawa}\ \emph {et~al.}(2019)\citenamefont
  {Terasawa}, \citenamefont {Matsumoto}, \citenamefont {Ozaki},\ and\
  \citenamefont {Gohda}}]{2019_Liechtenstein_Method_JPSJ}%
  \BibitemOpen
  \bibfield  {author} {\bibinfo {author} {\bibfnamefont {A.}~\bibnamefont
  {Terasawa}}, \bibinfo {author} {\bibfnamefont {M.}~\bibnamefont {Matsumoto}},
  \bibinfo {author} {\bibfnamefont {T.}~\bibnamefont {Ozaki}},\ and\ \bibinfo
  {author} {\bibfnamefont {Y.}~\bibnamefont {Gohda}},\ }\bibfield  {title}
  {\bibinfo {title} {Efficient algorithm based on Liechtenstein method for
  computing exchange coupling constants using localized basis set},\ }\href
  {https://doi.org/10.7566/JPSJ.88.114706} {\bibfield  {journal} {\bibinfo
  {journal} {Journal of the Physical Society of Japan}\ }\textbf {\bibinfo
  {volume} {88}},\ \bibinfo {pages} {114706} (\bibinfo {year}
  {2019})}\BibitemShut {NoStop}%
\bibitem [{\citenamefont {He}\ \emph {et~al.}(2021)\citenamefont {He},
  \citenamefont {Helbig}, \citenamefont {Verstraete},\ and\ \citenamefont
  {Bousquet}}]{2021_TB2J_J_Greens_fn_CPC}%
  \BibitemOpen
  \bibfield  {author} {\bibinfo {author} {\bibfnamefont {X.}~\bibnamefont
  {He}}, \bibinfo {author} {\bibfnamefont {N.}~\bibnamefont {Helbig}}, \bibinfo
  {author} {\bibfnamefont {M.~J.}\ \bibnamefont {Verstraete}},\ and\ \bibinfo
  {author} {\bibfnamefont {E.}~\bibnamefont {Bousquet}},\ }\bibfield  {title}
  {\bibinfo {title} {{TB2J}: A python package for computing magnetic
  interaction parameters},\ }\href {https://doi.org/10.1016/j.cpc.2021.107938}
  {\bibfield  {journal} {\bibinfo  {journal} {Computer Physics Communications}\
  }\textbf {\bibinfo {volume} {264}},\ \bibinfo {pages} {107938} (\bibinfo
  {year} {2021})}\BibitemShut {NoStop}%
\end{thebibliography}
\end{document}